\documentclass[useAMS,fleqn,usenatbib]{mnras}
\usepackage[T1]{fontenc}
\usepackage{ae,aecompl}

\usepackage{amsmath}
\usepackage{amsopn}
\usepackage[british]{babel}
\usepackage[varg]{txfonts}
\usepackage{biblio}
\usepackage{natbib}
\usepackage{color}

\usepackage{graphicx}
\usepackage{xpatch}
\makeatletter
\xpatchcmd\Gin@setfile{\expandafter\strip@prefix\meaning\@tempa}{\makebox[\Gin@req@width]{Image place holder (draft mode)}}{}{}
\makeatother
\usepackage{microtype}

\definecolor{orange}{rgb}{1.0,0.5,0.}

\def\MDM{\ifmmode{\>M_{\textnormal{\sc dm}}}\else{$$M_{\textnormal{\sc dm}}}\fi}

\def\XH{\ifmmode{\>X_{\textnormal{\sc h}}} \else{$X_{\textnormal{\sc h}}$}\fi}
\def\nH{\ifmmode{\>n_{\textnormal{\sc h}}} \else{$n_{\textnormal{\sc h}}$}\fi}

\def\maspyr{\ifmmode{\>\textnormal{mas~yr}^{-1}}\else{mas~yr$^{-1}$}\fi}

\def\mG{\ifmmode{\>\mu\mathrm{G}}\else{$\mu$G}\fi}
\def\erg{\ifmmode{\> {\rm erg}}\else{erg}\fi}
\def\keV{\ifmmode{\> {\rm keV}}\else{keV}\fi}

\def\deg{\ifmmode{\>^{\circ}}\else{$^{\circ}$}\fi}
\def\onedeg{\ifmmode{\>1^{\circ}}\else{$1^{\circ}$}\fi}

\def\xvir{\ifmmode{\>\!x_{vir}}\else{$x_{vir}$}\fi}
\def\Mvir{\ifmmode{\>\!M_{vir} }\else{$M_{vir} $}\fi}
\def\rvir{\ifmmode{\>\!r_{vir}}\else{$r_{vir}$}\fi}
\def\vvir{\ifmmode{\>\!v_{vir}}\else{$v_{vir}$}\fi}
\def\Vvir{\ifmmode{\>\!V_{vir} }\else{$V_{vir} $}\fi}

\def\tratio{\ifmmode{\>\tau}\else{$\tau$}\fi}

\def\rms{\ifmmode{\>r_{\textnormal{\sc ms}}}\else{$r_{\textnormal{\sc ms}}$}\fi}

\def\Mpc{\ifmmode{\>\!{\rm Mpc}} \else{Mpc}\fi}
\def\kpc{\ifmmode{\>\!{\rm kpc}} \else{kpc}\fi}
\def\pc{\ifmmode{\>\!{\rm pc}} \else{pc}\fi}

\def\Gyr{\ifmmode{\>\!{\rm Gyr}} \else{Gyr}\fi}
\def\Myr{\ifmmode{\>\!{\rm Myr}} \else{Myr}\fi}
\def\yr{\ifmmode{\>\!{\rm yr}} \else{yr}\fi}
\def\pyr{\ifmmode{\>\!{\rm yr}^{-1}}\else{yr $^{-1}$} \fi}
\def\s{\ifmmode{\>\!{\rm s}}\else{s}\fi}
\def\ps{\ifmmode{\>\!{\rm s}^{-1}}\else{s$^{-1}$}\fi}
\def\Hz{\ifmmode{\>\!{\rm Hz}}\else{Hz}\fi}

\def\kms{\ifmmode{\>\!{\rm km\,s}^{-1}}\else{km~s$^{-1}$}\fi}

\def\K{\ifmmode{\>\!{\rm K}}\else{K}\fi}

\def\sr{\ifmmode{\>\!{\rm sr}}\else{sr}\fi}
\def\psr{\ifmmode{\>\!{\rm sr}^{-1}}\else{sr$^{-1}$}\fi}
\def\arcs{\ifmmode{\>\!{\rm arcsec}}\else{arcsec}\fi}
\def\parcs{\ifmmode{\>\!{\rm arcsec}^{-1}}\else{arcsec${-1}$}\fi}
\def\parcss{\ifmmode{\>\!{\rm arcsec}^{-2}}\else{arcsec${-2}$}\fi}

\def\cm{\ifmmode{\>\!{\rm cm}}\else{cm}\fi}
\def\cc{\ifmmode{\>\!{\rm cm}^{3}}\else{cm$^{3}$}\fi}
\def\sqc{\ifmmode{\>\!{\rm cm}^{2}}\else{cm$^{2}$}\fi}
\def\pcc{\ifmmode{\>\!{\rm cm}^{-3}}\else{cm$^{-3}$}\fi}
\def\psc{\ifmmode{\>\!{\rm cm}^{-2}}\else{cm$^{-2}$}\fi}

\def\g{\ifmmode{\>\!{\rm g}}\else{g}\fi}
\def\Msun{\ifmmode{\>\!{\rm M}_{\odot}}\else{M$_{\odot}$}\fi}
\def\hMsun{\ifmmode{\> h^{-1}{\rm M}_{\odot}}\else{$h^{-1}$M$_{\odot}$}\fi}

\def\Zsun{\ifmmode{\>\!{\rm Z}_{\odot}}\else{Z$_{\odot}$}\fi}

\def\Lsun{\ifmmode{\>\!{\rm L}_{\odot}}\else{L$_{\odot}$}\fi}

\def\rayl{\ifmmode{\>\!{\rm R}}\else{R}\fi}
\def\mR{\ifmmode{\>\!{\rm mR}}\else{mR}\fi}

\renewcommand{\ion}[2]{\hbox{#1\,{\sc #2}}}

\def\lya{\ifmmode{\>\!{\rm Ly}\alpha}\else{Ly$\alpha$}\fi}

\def\Ha{\ifmmode{\>\!{\rm H}\alpha}\else{H$\alpha$}\fi}
\def\Hb{\ifmmode{\>\!{\rm H}\beta}\else{H$\beta$}\fi}

\def\HI{\ifmmode{\> \textnormal{\ion{H}{i}}} \else{\ion{H}{i}}\fi}
\def\HII{\ifmmode{\> \textnormal{\ion{H}{ii}}} \else{\ion{H}{ii}}\fi}
\def\CIV{\ifmmode{\> \textnormal{\ion{C}{iv}}} \else{\ion{C}{iv}}\fi}
\def\SiIV{\ifmmode{\> \textnormal{\ion{S}{iv}}} \else{\ion{Si}{iv}}\fi}

\def\NH{\ifmmode{\> {\rm N}_{\rm H}} \else{N$_{\rm H}$}\fi}
\def\Ng{\ifmmode{\> {\rm N}_{\rm gas}} \else{N$_{\rm gas}$}\fi}
\def\NHI{\ifmmode{\> {\rm N}_{\HI}} \else{N$_{\HI}$}\fi}
\def\MHI{\ifmmode{\> {\rm M}_{ \HI}} \else{M$_{\HI}$}\fi}

\def\mua{\ifmmode{\>\mu_{ \textnormal{\Ha}}}\else{$\mu_{ \textnormal{\Ha}}$}\fi}
\def\alphabha{\ifmmode{\>\alpha_{B}^{(\textnormal{\Ha})}}\else{$\alpha_{B}^{(\textnormal{\Ha})}$}\fi}

\newcommand{\myemail}{ttepperg@gmail.com}

\newcommand{\gaia}{{\em Gaia}}

\defcitealias{ast13a}{\textcolor{black}{Astropy Collaboration} 2013}
\defcitealias{ast18a}{\textcolor{black}{Astropy Collaboration} 2018}

\title[ The M31/M33 tidal interaction ]{ The M31/M33 tidal interaction: A hydrodynamic simulation of the extended gas distribution }

\author[Tepper-Garc\'\i{}a, Bland-Hawthorn, and Li]{%
Thor Tepper-Garc\'\i{}a,$^{1,2}$\thanks{\myemail} Joss Bland-Hawthorn,$^{1,2}$
and Di Li$^{3,4,5}$
\\
$^1$Sydney Institute for Astronomy, School of Physics, University of Sydney, NSW 2006, Australia\\
$^2$Centre of Excellence for All Sky Astrophysics in Three Dimensions (ASTRO-3D), Australia\\
$^3$Key Laboratory of the Five-hundred-meter Aperture Spherical radio Telescope (FAST),\\
National Astronomical Observatories, Chinese Academy of Sciences, Beijing 100101, China\\
$^4$School of Astronomy and Space Science, University of Chinese Academy of Sciences, Beijing 101408, China\\
$^5$NAOC-UKZN Computational Astrophysics Centre, University of KwaZulu-Natal, Durban 4000, South Africa
}

\date{Accepted 2020 January 30. Received 2020 January 30; in original form 2019 November 26}

\pubyear{\date{year}}

\begin{document}
\label{firstpage}
\pagerange{\pageref{firstpage}--\pageref{lastpage}}
\maketitle

\pdfminorversion=5
%--------------------------------------------------------------------------------------------------------------------------------------------------------------------------------
\begin{abstract}
% Word count: 239 (max. allowed by MNRAS: 250)
We revisit the orbital history of the Triangulum galaxy (M33) around the Andromeda galaxy (M31) in view of the recent \gaia\ Data Release 2 proper motion measurements for both Local Group galaxies. Earlier studies consider highly idealised dynamical friction, but neglect the effects of dynamical mass loss. We show the latter process to be important using mutually consistent orbit integration and $N$-body simulations. 
Following this approach we find an orbital solution that brings these galaxies to within $\sim 50$ kpc of each other in the past, $\sim 6.5$ Gyr ago.
We explore the implications of their interaction using an $N$-body/hydrodynamical simulation with a focus on the origin of two prominent features: 1) M31's Giant Stellar Stream; and 2) the M31-M33 \HI\ filament. We find that the tidal interaction does not produce a structure reminiscent of the stellar stream that survives up to the present day. In contrast, the M31-M33 \HI\ filament is likely a fossil structure dating back to the time of the ancient encounter between these galaxies.  Similarly, the observed outer disc warp in M33 may well be a relic of this past event. Our model suggests the presence of a tidally induced gas envelope around these galaxies, and the existence of a diffuse gas stream, the `Triangulum stream', stretching for tens of kpc from M33 away from M31. We anticipate upcoming observations with the recently commissioned, Five-hundred-meter Aperture Spherical radio Telescope (FAST) that will target the putative stream in its first years of operation.
\end{abstract}

%--------------------------------------------------------------------------------------------------------------------------------------------------------------------------------

%--------------------------------------------------------------------------------------------------------------------------------------------------------------------------------
\section{Introduction} \label{sec:intro}

The archaeology of the Local Group is of profound importance in modern astrophysics. \citet[][]{kou17a} show that the Local Group is very typical of galaxy groups in the local Universe since most are dominated by a few $L_\star$ galaxies. There are many questions we can ask about the Local Group that are presently impossible to consider for any other group. Within a few decades, we are likely to have 6D phase space coordinates for all major Local Group galaxies, including many of the constituents like globular clusters and dwarf galaxies. This may even allow us to `unravel' the orbits over billions of years to see how the Local Group has evolved dynamically.

M31 is comparable to or possibly larger than the Milky Way by mass \citep[e.g.][]{wat10a,dia14a,car17a,mcm17a}. Many recent observations serve to show key differences between both systems, due to the different accretion histories over the past 10 Gyr \citep[][]{kor13a,fer16a}. M31's bulge is more dominant than the compact bulge/bar system in the Milky Way \citep[][]{cou11a}; its central supermassive black hole is also 50 times more massive than Sgr A* \citep[][]{bla16a}. Long-running stellar surveys reveal a wealth of evidence for late major mergers throughout M31's halo (SPLASH: \citealt[][]{gil09b}; PHAT: \citealt[][]{dal12b}; PAndAS: \citealt[][]{iba14a}). Notably, the PAndAS survey has spent the last decade mapping out the environments of both M31 and M33 to reveal a rich structure of faint stellar wisps and streams \citep[][]{mcc09b,lew13b,mcc18b}.

The disk of M31 also shows major differences from the Milky Way: these include a very active star formation history  over the past 4 Gyr \citep[][]{ber15a,wil15a}, and a thick stellar disc that dominates over the thin disc component \citep[][]{dor15a}. The Milky Way's recent history is relatively quiescent, in particular, most of the stellar mass is locked up in a thin stellar disc.

The recent surveys of M31's environment have focussed on the stellar components with the goal to understand M31's assembly history, but there are certainly important clues about its accretion history from the gas phases as well. The limited neutral hydrogen (\HI) surveys to date have already revealed interesting structures, e.g., \citet[][]{bra04a} identified \HI\ between M31 and M33, part of which was further resolved to discrete structures by the GBT \citep[][]{wol13a} and some was shown to display a coherent structure \citep[][]{ker16a}. Gas streams around M31 have been detected in emission \citep[21 cm \HI; e.g.][]{loc12a} and in absorption \citep[ultra-violet atomic transitions; ][]{koc15a} extending from M31 both away from and towards M33. Similarly, the gas around M33 detected in the form of \HI\ displays a disturbed morphology \citep[][]{put09a}. We need to understand where to look near M31 and how we should target diffuse or clumpy gas. We can also look for gas associated with the stellar streams, and address open questions surrounding the known \HI\ features. For example, is the gas bridge observed between M31 and M33 the result of an interaction \citep[][]{bek08a} or rather the product of condensation within an intergalactic filament \citep[][]{wol13a,wol16a}? What is the origin of M33's disturbed \HI\ (and stellar) morphology?

New radio telescopes coming online, e.g., the Five-hundred-meter Aperture Spherical radio Telescope \citep[FAST;][]{nan11a,li16a}, will soon allow us to survey M31's and M33's gas distribution more extensively and to lower column density threshold, which can survive the cosmic and galactic ionising intensity out to 50-100 kpc in galactic radius if sufficiently clumpy \citep[][]{bla17a}. With the FAST L-band array of 19 feed-horns (FLAN), we have realised an unprecedented commensal survey mode for searching pulsars and imaging \HI\ simultaneously \citep[][]{li18a}, which enables a so-called deep Galactic-plane and Andromeda Survey (GAS). GAS will be sensitive to \HI\ column density as low as $5\times10^{16}$ \psc, a significant improvement over all previous studies.

Many of the issues surrounding how to interpret M31's complex environment come down to an incomplete understanding of how it has interacted with M33, its most massive companion, in the last few billion years. A close interaction ($< 100$ kpc) in the past could in principle generate tidal features that survive today \citep[e.g.][]{put09a,mcc09b}. However, a recent study based on ESA \gaia\ Data Release 2 (DR2) proper motions (PM) of the M31/M33 system appear to rule out this possibility \citep[][]{van19a}.
But this important study did not carry out a self-consistent dynamical simulation, i.e., they did {\it not} consider the problem of mass loss from either or both systems in their backwards integration. Whether two systems are on first infall or have experienced multiple orbits about each other cannot be reliably determined without consideration of mass loss \citep[][]{nic09a}. 

We revisit the orbital evolution of M31 and M33 with a more sophisticated treatment involving $N$-body simulations and forward-backward orbit integration, accounting for dynamical friction {\em and} dynamical mass loss in a self consistent way. Our work presents the first combined $N$-body and hydrodynamical (HD) simulation of the M31/M33 system to make use of the \gaia\ DR2 PM constraint. Earlier simulations of the tidal interaction between M31 and M33 \citep[e.g][]{bek08a,mcc09b} were constrained by (now outdated) M33's PM alone -- since M31's PM were not available at the time, or targeted specific aspects of the rich structure observed around in and around either M31 \citep[e.g.][]{ham18a} or M33 \citep[e.g.][]{sem18a}. Our particular focus is the distribution of the extended gas around {\em both} M31 and M33, with a view to understanding their gas accretion history, and to aid our upcoming observations with the FAST telescope. 

%--------------------------------------------------------------------------------------------------------------------------------------------------------------------------------
\section{Orbital history of the M31/M33 System} \label{sec:orb}

%--------------------------------------------------------------------------------------------------------------------------------------------------------------------------------
\subsection{Preliminaries} \label{sec:pre}

If one ignores their environment, the orbital history of two galaxies with respect to one another is governed by the following factors: 1) their relative position and velocity; 2) the mass distribution (i.e. the gravitational potential) of each galaxy; 3) the effect of dynamical friction (DF) of the galaxies action upon one another; 4) their mass evolution, i.e. mass loss due to tidal stripping, generally experienced by the smallest object. A widespread approach in inferring the orbital history of two galaxies with respect to one other is to take its present-day orbital parameters, and integrate the equations of motion backwards in time. In doing so, each galaxy is approximated by an extended body with a prescribed mass distribution. We refer to this approach as 'semi-analytic orbit integration' (SAOI).

This approach is a strict {\em two-body} calculation, and it therefore necessarily ignores the effects of dynamical friction and dynamical mass loss (a.k.a. tidal stripping). These need to be accounted for using idealised parameterisations which themselves rely on a number of assumptions. A survey of the relevant literature reveals that, quite surprisingly, it appears to be a common practice to ignore the effect of tidal stripping altogether, with a few notable exceptions \citep[e.g.][]{jia00a,nic09a,die17a}.

This is true for the M31/M33 system as well. The common orbital history of these galaxies has been repeatedly studied in the past in response to improvements in the measurement of their PM \citep[e.g.][]{van12d,sal16a,pat17b,sem18a}. Most recently, \citet[][]{van19a} have explored the implications of the new \gaia\ DR2 PM for these galaxies. They concluded, as did \citet[][]{van12d} and \citet[][]{pat17b} before them using different PM data sets, that M33 is likely on its first approach to M31. However, their calculations ignore the effect of tidal stripping (as did other before them). These authors include the effect of DF in their analytic calculations using the calibration with $N$-body simulations obtained by \citet[][]{van12d}. Interestingly, the latter find that the semi-analytic M31/M33 orbit (projected into the future) decays more rapidly compared to their $N$-body result. This is likely caused by the absence of tidal stripping in their semi-analytic calculations.

We calculate anew the orbital history of M33 around M31 accounting for the effect of dynamical friction {\em and} mass loss due to tidal stripping in a self-consistent way. To this end, we choose to use the \gaia\ DR2 PM alone \citep[and not, say, the weighted \gaia\ / {\em Hubble Space Telescope} measurements; see][]{van19a} since they represent the most extreme of the PM measurements in the sense that they have been claimed to {\em exclusively} support orbital histories where M33 is on its first approach to M31 \citep[][]{van19a}. At the other extreme are the PM measurements by \citet[][]{sal16a}, which allow orbital histories with at least one close encounter between these galaxies with pericentric distances well below 100 kpc \citep[e.g.][]{sem18a}.

Note that we ignore throughout the presence of additional members of the Andromeda group as well as the Milky Way and its satellites (e.g., the Magellanic Clouds) and their gravitational perturbation of the M31/M33 orbit. As shown by \citet[][]{pat17b}, this is a valid approximation.

%--------------------------------------------------------------------------------------------------------------------------------------------------------------------------------
\subsection{Semi-analytic orbit integration} \label{sec:saoi}

Our SAOI approach is as follows.\footnote{Our integration code is freely available upon request via email to the corresponding author (TTG).}
The dynamics of the two-body system is governed by the following coupled 3D equation of motions:
\begin{align}
	\ddot{\vec{r}}_{31}  & = -\vec{\nabla} \phi_{33} \, , \label{eq:eom1}\\
	\ddot{\vec{r}}_{33}  & = -\vec{\nabla} \phi_{31} + \vec{f}_{\textnormal{\sc df}} \, , \label{eq:eom2}
\end{align}
where $r \equiv |\vec{r}_{33} - \vec{r}_{31}|$ is the relative distance between M31 and M33; and $\vec{f}_{\textnormal{\sc df}}$ is the acceleration due to dynamical friction, discussed below. The gravitational field is given by
$-\vec{\nabla} \phi_{31,33}(r) = -( G M_{31,33}(r) / r^2 ) \hat{r} \, .$
Here, $G$ is the gravitational constant, $\hat{r}$ is the unit vector in the direction of $\vec{r}$, and $M_{31,33}(r)$ describe the mass distributions of M31 and M33  within $r$.

To obtain an orbital solution, we integrate the system of equations (\ref{eq:eom1} - \ref{eq:eom2}) backwards in time for roughly 8 Gyr using a leapfrog ('kick-drift-kick') scheme. This scheme is fully time-reversible, and thus appropriate to ensure that the backwards and forward integration of the orbit are consistent with one another with a fixed time step of $\Delta t = 10^{-3}$ Gyr, accounting for the effect of dynamical friction and mass loss due to tidal stripping as described below. Note that our adopted time step size is small enough to avoid energy drift.

Our orbital solution will depend on a number of factors: 1) the value of the orbital initial conditions; 2) the mass model of each galaxy; 3) the prescription to calculate the effect of dynamical friction; and 4) the prescription to account for the effect of tidal stripping. We describe each of these next.

%--------------------------------------------------------------------------------------------------------------------------------------------------------------------------------
\subsubsection{Orbital initial conditions}  \label{sec:oic}

The orbital initial conditions are set by the following present-day orbital parameters. The position of M33 relative to M31 is \citep[][]{van12c,sem18a}
\begin{equation} \label{eq:r0}
	\vec{r}_0 = (-97 \pm 23, -122 \pm 35, -130 \pm 19) ~\kpc \, ,
\end{equation}
corresponding to a present-day relative distance $d_0 = 203 \pm 27$ kpc. The total uncertainty is calculated assuming simple error propagation and uncorrelated measurement error among the coordinates. We adopt the central value of the velocity from the \gaia\ DR2 proper motion (PM) measurements alone \citep[][]{van19a}, 
$\vec{v}_{0;31} = (0\pm75, 176\pm51, -84\pm73) ~\kms$ 
and 
$\vec{v}_{0;33} = (49\pm74, 14\pm70, 28\pm73)~\kms$, 
which yield a relative velocity of
\begin{equation} \label{eq:v0}
	\vec{v}_0 = (49 \pm 105,190 \pm 87,112 \pm 104) ~\kms \, ,
\end{equation}
corresponding to a present-day relative space velocity of $v_0 = 226 \pm 92$ \kms, with a radial component of -209 \kms, and a tangential component of 85  \kms. The uncertainty on these PM measurements is still large, but consistent within $1 \sigma$ with earlier measurements taken with the {\em Hubble Space Telescope} \citep[HST;][]{van12d}. All of the above given values can be found in Tab.~\ref{tab:orbit}.

%--------------------------------------------------------------------------------------------------------------------------------------------------------------------------------
\subsubsection{Gravitational potential}  \label{sec:pot}

For the purpose of our SAOI, we approximate M31 and M33 each by a spherical DM halo, and ignore for now other components (stellar bulge, stellar disc, gas disc, hot halo). Since the mass of each galaxy is dominated by its host DM halo, and the shortest relative distance between the galaxies is well beyond their optical radii along any plausible orbit, this approximation does not affect our orbital integration. Indeed, as we show in the next section, the orbital history we find with our SAOI approach agrees well with the corresponding result obtained from a pure $N$-body simulation -- which does account for the full multi-component nature of the two galaxies, thus indicating that the error introduced by our approximation is negligible. It is perhaps important to note that the total mass and its distribution both in M31 and M33 are still highly uncertain. This allows us to choose suitable parameter values within a broad range consistent with observations.

Following \citet[][]{pat17b}, M31's DM host halo is assumed to be well described by a \citet[][NFW]{nav97a} profile; the DM halo of M33 is approximated by a \citet[][]{plu11a} sphere. Each galaxy model is then fully specified by its total mass $M_t$, a scale radius $r_s$, and a truncation radius $r_{tr}$ (see Tab.~\ref{tab:comp}). The total mass of our M31 model is $\sim 2 \times 10^{12}$ \Msun\ and of our M33 model, $\sim 1.6\times10^{11}$ \Msun.

The mass of M31 is consistent with the upper limit of the mass of this galaxy based on limits to the total mass of the Local Group \citep[$\sim 3 \times 10^{12} \Msun$; e.g.][see also \citealt{far13a}]{dia14a}, and independent estimates of the mass of the Milky Way \citep[$\sim 1 \times 10^{12}$ \Msun; e.g.][]{mcm17a}.

The mass of M33 we adopt is lower than traditional estimates based on the modelling of its observed rotation curve. Using this approach, \citet[][]{kam17b} estimate a virial mass for M33 of $\sim 5 \times 10^{11}$ \Msun{} \citep[see also][]{cor14b} . As discussed by these authors, however, with a baryonic mass of order $\lesssim 10^{10}$ \Msun, the baryonic:total mass fraction in M33 is only $\lesssim 2$ percent, well below the typical value for disc galaxies \citep[$\sim 7$ percent; e.g.][]{zar14a}. Estimates of the mass of a galaxy based on the rotation curve alone out to radii beyond the maximum observed extension are highly model dependent. If instead we assume the estimates of the total baryonic fraction to be more robust, then the average baryon:total mass fraction in disc galaxies implies a total mass for M33 at the present-epoch of $\sim 1.5 \times 10^{11}$ \Msun, consistent with the value we adopt.

%--------------------------------------------------------------------------------------------------------------------------------------------------------------------------------
\subsubsection{Dynamical friction}  \label{sec:df}

Given its larger mass, M31's DM halo is likely to be more extended than M33's at all times. Therefore, we assume that M33 experiences the effect of dynamical friction by the DM matter field of M31 but not vice-versa. This assumption is reflected in the form of equations (\ref{eq:eom1}) and (\ref{eq:eom2}). Following previous work, the acceleration due to dynamical friction is accounted for using a special case derived from the generic Chandrasekhar formula:
\begin{equation} \label{eq:df}
	\vec{f}_{\textnormal{\sc df}} = - 4 \pi G^2 \ln \Lambda ~\rho^{\textnormal{\sc dm}}_{31} M_{33} ~\left\{ {\rm erf}\left[\chi \right] - \frac{2 \chi}{\sqrt{\pi}} \exp \left[-\chi^2 \right] \right\} ~\frac{\vec{v}}{v^3} \, , 
\end{equation}
where $\rho^{\textnormal{\sc dm}}_{31}(r)$ describes M31's DM density field; $v \equiv |\vec{v}|$ is the relative velocity of M31 and M33; erf$[\cdot]$ is the error function, $\chi \equiv \ v / \sqrt{2} \sigma^{\textnormal{\sc dm}}(r)$; here, $\sigma^{\textnormal{\sc dm}}(r)$ is the one-dimensional velocity dispersion of M31's DM matter field. The term $\ln \Lambda $ is the Coulomb logarithm, discussed below.

It is worth keeping in mind the assumptions underlying equation \eqref{eq:df}. First, it is assumed that the mass of the DM particle is negligible compared to M33's total mass at any point along the orbit, i.e.  $M_{33}(r) \gg m_{\textnormal{\sc dm}}$. Second, it is assumed that the velocities of the DM particles obey a Maxwellian distribution. While the former may seem reasonable, the latter may in fact not be justified in general.

The velocity dispersion is calculated using the analytic fit by \citet[][]{zen03a}:
\begin{equation} \label{eq:sig}
	\sigma^{\textnormal{\sc dm}}(r) = v_{max} \frac{1.4393 ~x^{0.354}}{1 + 1.1756 ~x^{0.725}} \, ,
\end{equation}
appropriate for an NFW profile and accurate to 1 percent in the range $x = [10^{-2}, 10^2]$. Here, $x = r / r_s$ and $v_{max}$ is the peak value of the circular velocity of the DM halo,\footnote{In practice, we use a root finding algorithm to obtain $v_{max}$ from the expression for the NFW circular velocity,
\begin{equation}  \label{eq:vnfw} \notag
	v_c^2(r) = 4 \pi G \rho_0 r_s^2  \left[ \frac{\ln (1 + x)}{x}  - \frac{1}{1 + x} \right] \, ,
\end{equation}
where $\rho_0$ is the density scale of the corresponding NFW profile.
} which for an NFW profile is found at a distance from the centre $r \approx 2.16 r_s$.

The Coulomb logarithm is a measure of the scattering between the object experiencing dynamical friction and the background matter field responsible for the latter. The Coulomb logarithm is, in fact, unknown in most cases of interest in galactic dynamics, and a number of parameterisations exist. We adopt \citet[][]{has03a}'s,
$\ln \Lambda = r / 1.4 \epsilon$.
Here, $\epsilon$ is the so-called `softening length', usually taken to be the (equivalent) Plummer scale radius of the mass model. It is a free, and thus a tuneable parameter. Its value can only be guessed, and it is best set via a trial-and-error approach in an iterative way comparing the outcome of the SAOI and a corresponding $N$-body simulation. We note that other authors (see Sec. \ref{sec:orb}) adopt slightly different values for the constant factor in the denominator. This is irrelevant since it is the product of this factor and $\epsilon$ that determines the behaviour of this particular parametrisation of $\ln \Lambda $. As discussed in the next section, the value that leads to the best match between our SAOI and an $N$-body simulation is $\epsilon \approx 3.6$ kpc. It is worth noting that, somewhat ironically, the latter value is roughly a fraction 1/5 of the value of the Plummer scale radius we use for M33's mass model (see Tab.~\ref{tab:comp}).

%--------------------------------------------------------------------------------------------------------------------------------------------------------------------------------
\subsubsection{Dynamical mass loss}  \label{sec:dml}

As mentioned earlier, a key difference between our approach and similar previous work is that we take into account the effect of mass loss due to tidal stripping. We assume that only M33 experiences the effect of stripping by the tidal field of M31. In other words, we assume that M31's mass is constant during the orbital evolution of the two galaxies, and it does not require any special treatment during the approach outlined below. This also implies that we neglect M31's {\em cosmological} mass evolution. This approximation should, however, not affect our results significantly. A recent statistical analysis of mass assembly  in large-scale cosmological simulations \citep[][]{san20a} suggest that M31-like galaxies had typically acquired well above 50 percent of their present-day mass by the time M33 experiences a close encounter with M31 in our new orbit (see Sec.~\ref{sec:oh}).

%--------------------------------------------------------------------------------------------------------------------------------------------------------------------------------
% Table: Orbital parameters
\begin{table*}
\begin{center}
\caption{Infall- and present-day parameters. All values correspond to properties of M33 relative to M31. In all cases the relative position at infall ($\sim 8$ Gyr in the past) is $\vec{r}_{-8} = (-321,306,-172)$ kpc, corresponding to a relative distance of 475.7 kpc, and the relative velocity at infall is $\vec{v}_{-8} = (175,-107,116)$ \kms, corresponding to a relative speed 235.3 \kms. Position and distances in kpc; velocities and speeds in \kms. No errors given for results other than those obtained from observations.}
\label{tab:orbit}
\begin{tabular}{lcccc}
\hline
\hline
		 	&  Position $\vec{r}_0 = (x,y,z)$					& Distance		& Velocity $\vec{v}_0 = (v_x,v_y,v_z)$ 			& Speed\\
\hline
Observations \\
			& $(-97 \pm 24, -122 \pm 35, -130 \pm 19)$ 		& $203\pm27$	& $(49 \pm 105,190 \pm 87,112 \pm 104)$  	& $226 \pm 92$\\
\hline
SAOI\\
			& (-86,-133,-124) 							& 201			& (37,190,101)								& 219 \\
\hline
$N$-body\\
			& (-66,-163,-140) 							& 224			& (4,186 ,71) 								& 199 \\
\hline
$N$-body/HD\\
			& (-64,-139,-126)							& 198			& (14, 201, 83)								& 218 \\
\hline
\end{tabular}
\end{center}
\end{table*}
%--------------------------------------------------------------------------------------------------------------------------------------------------------------------------------

Following \citet[][]{nic09a}, tidal stripping is implemented as follows. First, we calculate the tidal radius $r_t$ of M33 at each time step via \citep[][]{kly99a}
\begin{equation} \label{eq:tid}
	\left(\frac{r}{r_t}\right)^3 \frac{M_{33}(r_t)}{M_{31}(r)}  = 2 - \frac{r}{M_{31}(r)}\frac{dM_{31}(r)}{dr} \, .
\end{equation}
Usually, one assumes that the mass beyond the tidal radius is lost at any point along the orbit, in particular at each pericentric passage (because the tidal radius is smallest there).\footnote{The tidal radius is implicitly defined in equation \eqref{eq:tid} and it needs to be solved for with a root finding algorithm. We use the Brent method.} But this is a valid assumption only when moving {\em forward} in time along the orbit. A critical aspect in our approach is how to handle the effect of tidal stripping when integrating {\em backwards} in time. The inverted time arrow implies that mass has to {\em increase}. This is indeed a non-trivial task because calculating the bound mass at each pericentric passage recursively requires knowledge of the earlier (in time) bound mass (or the earlier pericentric distance), which is impossible unless the full orbit is known in advance.

Our strategy to tackle this difficulty is as follows. We assume an arbitrary reference mass and a present-day (initial) bound mass for M33. We integrate the orbit until we identify the first pericentric passage (if any) along the orbit, and calculate there the tidal radius and the corresponding bound mass of M33 using the reference mass. The new bound mass is then set to the maximum of the previous and the new bound masses, which ensures that M33's mass is a monotonically increasing function along the backwards orbit.

The previous steps are repeated until the integration time is reached. The last bound mass obtained defines M33's infall mass.\footnote{In this context, `infall' simply refers to the epoch at which we start the calculation of the forward orbit, i.e., at $\sim$8 Gyr in the past. A more physically motivated definition could refer to the epoch at which M33 first crosses M31's virial radius. However, this distinction is irrelevant here.} Its value is used together with its corresponding orbital parameters as input for a forward integration. When integrating forward in time, we ensure that whenever the tidal radius increases (as a result of an increased relative distance between the galaxies), the bound mass does not increase. In other words, M33's mass is a strictly monotonically decreasing function along the forward orbit -- i.e., the inverse behaviour of M33's mass when integrating backwards.

We note that the use of a reference mass renders our approach not fully self-consistent at first because, during a backwards integration, the potential of M33 is defined by the reference mass, while the mass that enters into the calculation of the dynamical friction force is given by the bound mass at that point along the orbit. To guarantee self-consistency, one needs to make sure that the final (i.e., present-day) mass resulting from the forward integration matches (as closely as possible) the initial bound mass used for the backwards integration. Consistency with observations dictates the initial bound mass be given by the value of M33's present-day mass. The reference mass is thus the only free parameter in our approach to account for dynamical mass loss. The sole constraint on it is that it must be larger than the present-day mass of M33. It is important to realise that the reference mass is simply an auxiliary quantity that has no physical meaning and that plays no role in the final orbital solution once this has been found. In particular, the reference mass does not enter the calculation of the forward orbit.

Following a simple trial-and-error approach, and with some intuition, it has taken us on the order of 10 attempts to find the value of the reference mass such that the value of the infall and the final bound masses of M33 obtained during the backwards integration are consistent with their corresponding values obtained during the corresponding forward integration.\footnote{This procedure can be automated using an optimisation algorithm. We defer this task to a future work. \label{foo:aut}}

We find that a reference mass for M33 of $3.55\times10^{11}$ \Msun\ yields a present-day mass (at $\tau = 0$ Gyr) of $1.31\times10^{11}$ \Msun, and an infall mass (at $\tau \approx -8$ Gyr) of $2.31 \times10^{11}$ \Msun. Integrating forward (starting at $\tau \approx -8$ Gyr) adopting an infall mass of $2.63 \times10^{11}$ \Msun\ results in present-dat mass (at $\tau = 0$ Gyr) of $1.56\times10^{11}$ \Msun. It is worth emphasising that there is no need for a reference mass when integrating forward. The infall masses in both cases are broadly consistent with each other. Most importantly, the infall mass in the forward integration is consistent with the initial M33 mass of our $N$-body simulation. Similarly, the present-day mass in either the forward or the backward integration is consistent with the our estimate for the present-day mass of M33 ($\sim1.5\times10^{11}$ \Msun; see above).

Using these values together with our best value for $\epsilon$, a backwards integration of the M31/M33 system by roughly $8$ Gyr into the past constrained by their present-day orbital parameters (eqs. \ref{eq:r0} and \ref{eq:v0}) yields a relative infall position between the galaxies $\vec{r}_{-8} = (-321,306,-172)$ kpc, corresponding to an infall relative distance of 475.7 kpc, and a relative infall velocity $\vec{v}_{-8} = (175,-107,116)$ \kms, corresponding to an infall relative speed 235.3 \kms.

We stress that the values of the reference mass and $\epsilon$ are particular to the orbital parameters and to the galaxy mass distributions considered here, and there is no guarantee that, nor an obvious reason why, these values should be valid in general.

%--------------------------------------------------------------------------------------------------------------------------------------------------------------------------------
% FIGURE: Orbital history
\begin{figure*}
\centering
\includegraphics[width=0.5\textwidth]{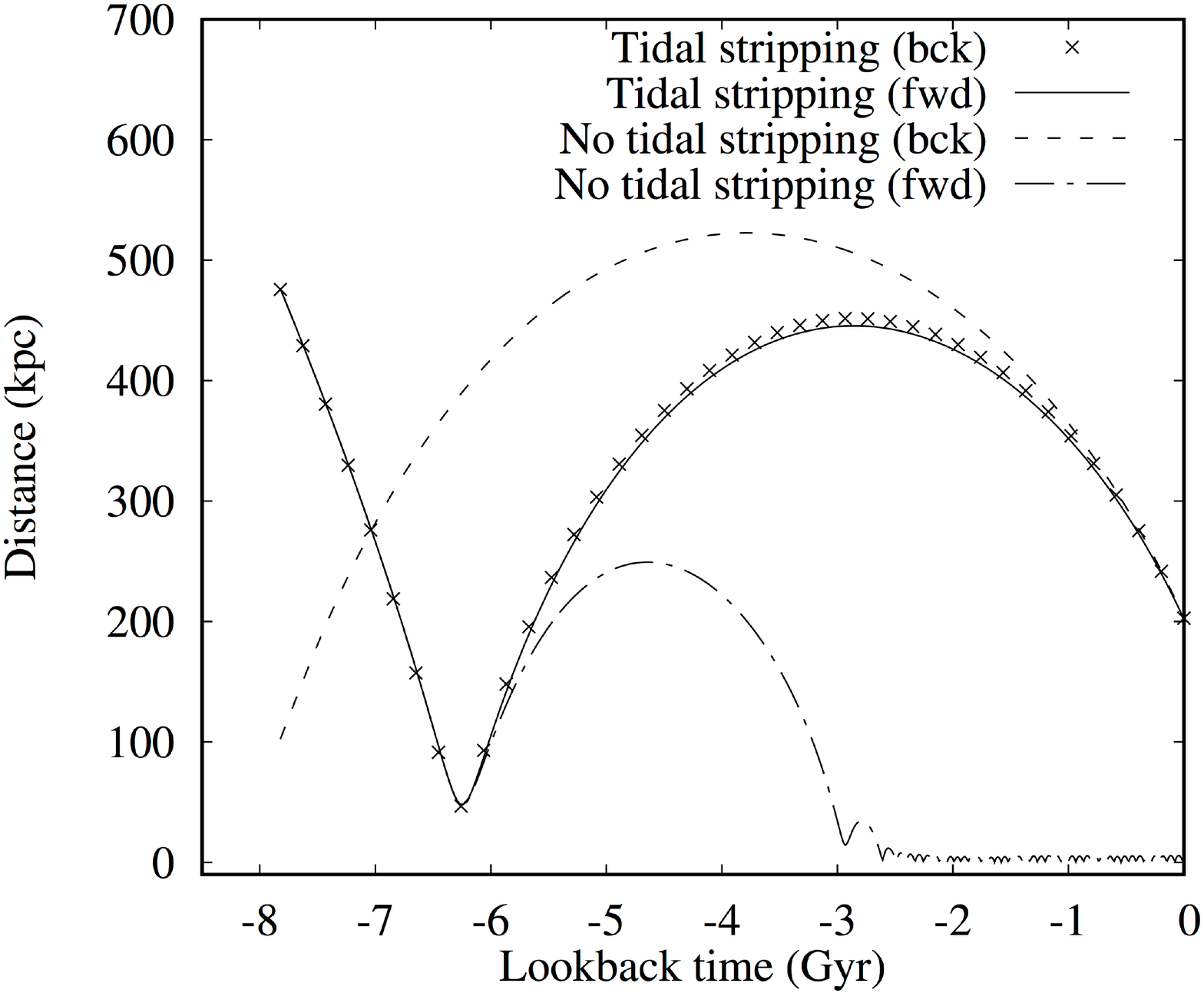}\hfill
\includegraphics[width=0.5\textwidth]{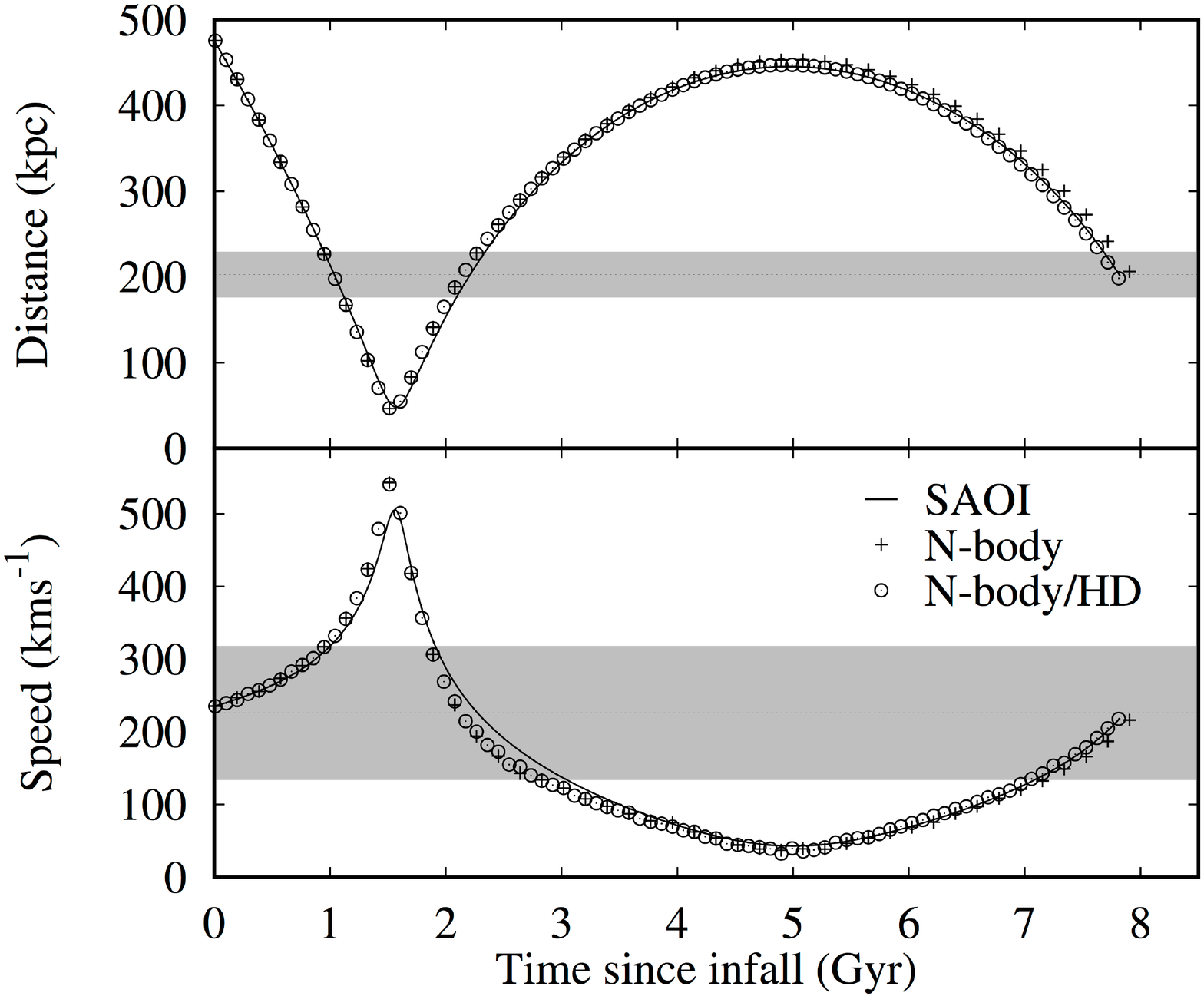}
\caption[ Orbital history ]{ Orbital history of the M31/M33 system. Left: A comparison of the orbital history obtained with a backwards (`bck') or forward (`fwd') SAOI with tidal stripping (referred to as `second-infall') or without it (`first-infall'). Note that the solid curve overlaps with the dot-dashed curve from infall (at $\sim 8$ Gyr in the past) up to the pericentric passage (at $\tau \approx 6.5$ Gyr). The infall epoch and the present-day are flagged by a look-back time of $\tau \approx -8$ Gyr and $\tau = 0$ Gyr, respectively. Right: Relative distance (top) and relative speed (bottom) between M33 and M31 in a second-infall scenario. The solid curves correspond to the result of our forward SAOI calculations. The plus signs (+) and the open circles ($\odot$) correspond to the distance (top) and speed (bottom) between M31 and M33 in an $N$-body simulation and a combined $N$-body/HD simulation, respectively. The gray-shaded area indicates either the observed relative distance or the observed relative speed and its associated uncertainty. In this panel, the infall epoch and the present-day are indicated by $t = 0$ Gyr and $t \approx 8$ Gyr, respectively. Note that the orbital history indicated by the solid curve in the left panel is identical to that indicated by the solid curve in the top sub-panel on the right. }
\label{fig:orbit}
\end{figure*}
%
%--------------------------------------------------------------------------------------------------------------------------------------------------------------------------------

%--------------------------------------------------------------------------------------------------------------------------------------------------------------------------------
\subsubsection{Orbital history}  \label{sec:oh}

The results from our SAOI calculation just described are displayed in Fig.~\ref{fig:orbit}. The left panel shows the orbital histories that result from a backwards integration accounting for the effect of tidal stripping (crosses; $\times$; henceforth referred to as `second infall') or without it (dashed curve). The latter result -- which corresponds to a 'first-infall' scenario, is virtually identical to the orbital history calculated by \citet[][their `high-mass M31/high-mass M33' case]{pat17b}, and serves as a cross-check for our calculations.

The right panel of Fig.~\ref{fig:orbit} displays the distance (top) and the  speed (bottom) between M31 and M33 along their common orbit in our second-infall scenario. The observed relative distance and the observed relative speed are indicated in each case by the horizontal dashed line and the shaded area. The full 3D information is displayed in Tab.~\ref{tab:orbit}. Both the relative distance and the relative speed resulting from our SAOI calculation match their observed counterpart well within their corresponding uncertainty.

Our time integration (leapfrog) scheme is time-reversible by construction, even accounting for the effect of dynamical friction. Our prescription to account for mass loss via tidal stripping is it as well to a good approximation. In the left panel of Fig.~\ref{fig:orbit} we visually compare the result from our forward (solid curve) and backward (crosses; $\times$) calculations. The agreement between these indicates that our backwards and forward orbital calculations are consistent with each other. It also shows that the reference (auxiliary) mass has no effect on the orbital solution, as claimed above.

The striking difference between the backwards orbit indicated by the crosses (second infall) and the backwards orbit indicated by the dashed curve (first infall) displayed in the left panel of Fig.~\ref{fig:orbit} demonstrates that the effect of mass loss due to tidal stripping significantly changes the nature of the orbital solution. This result is the interplay between two factors. First, the initial (i.e., present-day) M33 mass of the first-infall orbit is higher ($\sim 2.6 \times 10^{11}$ \Msun) compared to initial mass used in the second infall orbit ($\sim 1.3 \times 10^{11}$ \Msun). This is necessary for consistency. It ensures that the infall mass in both cases is roughly equal (i.e., $\sim 2.6\times10^{11}$ \Msun). Second, the fact that a reference mass ($3.55\times10^{11}$ \Msun, higher than either of the initial masses) is needed when tidal stripping is accounted for in a backwards integration.\footnote{The initial mass and reference mass are one and the same when tidal stripping is neglected, of course.}  As already stated (Sec.~\ref{sec:dml}), the initial mass affects the behaviour of the dynamical friction, but the effective gravitational attraction between the galaxies is determined by the reference mass. Therefore, a higher initial mass leads to a stronger `repulsion' when integrating backwards. At the same time, a higher reference mass yields a stronger pull between the galaxies. In consequence, the galaxies reach a larger separation over a given time span when tidal stripping is neglected, which results in the orbit without a close encounter (first infall). We stress that the inconsistency between which mass dominates which effect when accounting for dynamical mass loss becomes irrelevant {\em once an orbital solution has been found}, provided the orbit as well as the infall and present-day masses in a backwards integration and a forward integration are consistent with one another, as is the case here.

The strong effect of tidal stripping on the orbital evolution is clearly manifested by the forward integration. The dot-dashed curve in the left panel of Fig.~\ref{fig:orbit} shows the orbit resulting from identical initial conditions (infall mass, orbital parameters) as used in the second-infall solution (solid curve), but ignoring the effect of tidal stripping. These orbital histories are identical up to a few million years after the pericentric passage. As a result of tidal stripping then, M33 loses roughly half its mass, and the effect of DF becomes smaller, thus allowing the galaxy to reach a greater distance from M31. In contrast, ignoring mass loss due to tidal stripping leads to a stronger DF effect and thus a stronger orbital decay.  So strong in fact that M33 undergoes a number of additional pericentric passages before merging with M31.

It is worth noting that orbital solutions of the second-infall type with close encounters ($< 100$ kpc) are not new. \citet[][]{pat17b} make use of the {\em HST} PM -- which are consistent with the \gaia\ PM \citep[][]{van19a}, and find orbital solutions with typical pericentric distances of $\sim$100 kpc, and as low as $\sim$50 kpc. \citet[][]{sem18a} find orbital solutions with pericentric distances even below 50 kpc, but their results are based on orbital parameters which are not consistent with the \gaia\ DR2 PM measurements. However, our {\em self-consistent} orbital solution with a pericentric passage below 50 kpc based on the \gaia\ DR2 PM alone is a previously unknown result.

In summary, with a full account of the relevant physical processes at play (gravity, dynamical friction, tidal stripping) we have found a new orbit along which M33 underwent a close ($\lesssim 50$ kpc) pericentric passage from M31's centre some 6.5 Gyr ago. The question then arises of how this close encounter affected the evolution of these galaxies, and whether there are fossil imprints of this interaction observable today. In order to address this question, we resort to a combined $N$-body/HD simulation of the infall of M33 into M31.

%--------------------------------------------------------------------------------------------------------------------------------------------------------------------------------
\section{The infall of M33 onto M31} \label{sec:sims}

Prior to exploring the consequences that the orbital history we have found, we need to ensure we can recover our SAOI solution with a combined $N$-body/HD simulation. This is a necessary step given the simplifying assumptions underlying our SAOI calculation, in particular the idealised form of the dynamical friction and the prescription to account for dynamical mass loss. For now we opt to address this in an iterative way, as described below.\footnote{See Footnote \ref{foo:aut}.} Because $N$-body/HD simulations are expensive, we choose to use first pure $N$-body simulations.

%--------------------------------------------------------------------------------------------------------------------------------------------------------------------------------
\subsection{A minimal $N$-body model} \label{sec:nbo}

Dynamical friction and tidal stripping are naturally and self-consistently accounted for in $N$-body simulations, provided $N$ is `large' enough. Thus, for a given set of initial orbital parameters the most important factors determining the outcome of an $N$-body simulation of the infall of M33 into M31 is the mass model used to approximate each of the galaxies.

In contrast to our SAOI, we now assume that both M31 and M33 are {\em multi-component} systems each consisting a stellar disc embedded within a host DM halo. In addition, M31 is assumed to host a stellar bulge, as observed \citep[q.v.][]{mou13a}. The adopted profile and structural parameters for each of the components of each galaxy are given in Tab.~\ref{tab:comp}. Note the we ignore the gaseous components for now. These will be taken into account in our combined $N$-body/HD simulation discussed in the next section.

Other physical factors that likely affect the outcome of an $N$-body simulation include the initial orientation of the galaxy spins relative to their orbit. Following \citet[][]{sem18a}, M31's spin is initially roughly pointing along the $z$-axis; the spin of M33 is initially pointing the negative $x$-axis.

%--------------------------------------------------------------------------------------------------------------------------------------------------------------------------------
%TABLE: Component Parameters
\begin{table}
\begin{center}
\caption{Relevant model parameters (initial values).  Column headers are as follows: $M_t$ := total mass ($10^{9}$ \Msun); $r_s$ := scalelength (kpc); $r_{tr}$ :=  truncation radius (kpc); $N$ := particle number ($10^{5}$); $Z$ := gas metallicity ($\Zsun$).}
\label{tab:comp}
\begin{tabular}{lcccccc}
\hline
\hline
 					& Profile		& $M_t$ 		&  $r_s$ 		& $r_{tr}$ 	& $N$  & $Z$ ~\\
\hline
%M31\\
M31\\
\hline
DM halo				& NFW		& 1910		& 11.7		& 326	& 	10	&	--	~\\
Stellar bulge$^{\,a}$		& H			& 11.2		& 1			& 4		&	1	&	--	~\\
Stellar disc$^{\,a}$		& Exp		& 80.7		& 7.3$^{c}$	& 40		&	10	&	--	~\\
Hot halo		 		& NFW		& 3.10		& 11.7		& 326	&	--	&	0.3	~\\
Gas disc				& Exp$^{\,b}$	& 7.2	5		& 5.0	$^{d}$	& 60		&	--	&	0.3	~\\
\hline
M33\\
\hline
DM halo				& Plu		& 282		& 17.9		& 197	&	10	&	--	~\\
Stellar disc$^{\,a, \,f}$	& Exp		& 5.8			& 2.5$^{e}$	& 25		&	5	&	--	~\\
Gas disc				& Exp$^{\,b}$	& 5.8			& 2.5			& 40		&	--	&	0.2	~\\
\end{tabular}
\end{center}
\begin{list}{}{}
\item {\em Notes}. NFW, \citet[][]{nav97a} profile; H, \citet{her90a} profile; Plu, \citet[][]{plu11a} profile; Exp, Radial exponential profile

$^{a\,}$The stellar metallicity is ignored as it is of no relevance for our study.

$^{b\,}$In vertical hydrostatic equilibrium initially at $T = 10^4 ~\K$

$^{c\,}$Scaleheight set to 1 kpc.

$^{d\,}$Scaleheight set by vertical hydrostatic equilibrium (`flaring' disc).

$^{e\,}$Scaleheight set to 0.5 kpc.

$^{f\,}$Toomre's $Q > 1.8$ everywhere initially \citep[][]{ten17a}

 \end{list}
\end{table}
%--------------------------------------------------------------------------------------------------------------------------------------------------------------------------------

To follow the simplest approach possible, we consider $\epsilon$ to be the only free parameter of our SAOI calculation. With this assumption, we proceed to verify our semi-analytic as follows. First, we obtain the infall orbital parameters obtained from a backwards integration adopting a first guess for the value of $\epsilon$. A common choice is to set $\epsilon$ to the (equivalent) Plummer scale length of M33's mass model (see Tab.~\ref{tab:comp}). Then, we use these orbital parameters and an $N$-body representation of M31 and M33 to run a simulation forward in time. We then compare the orbital history of M33 relative to M31 obtained in the simulation to the SAOI result. In the likely event that no good agreement is found, we tune the value of $\epsilon$ and repeat the previous steps until a good match between our SAOI calculation and the $N$-body simulation is obtained. Our best value thus obtained is $\epsilon \approx 3.6$ kpc. The infall orbital parameters obtained from the backwards integration at $\sim 8$ Gyr in the past are  $\vec{r}_{-8} = (-321,306,-172)$ kpc, and $\vec{v}_{-8} = (175,-107,116)$ \kms.

We generate the pure $N$-body and the $N$-body/gas representations for each galaxy with the {\sc dice} code \citep[][]{per14c}. The corresponding initial conditions -- both the pure $N$-body and the $N$-body/hydrodynamic run -- are evolved with the {\sc Ramses} code \citep[version 3.0 of the code last described by][]{tey02a}. The limiting spatial resolution in our simulations is $\gtrsim 60$ pc at all times. The mass resolution for each component can be easily obtained from Tab.~\ref{tab:comp}, columns 3 and 6.\footnote{The setup files used to create initial conditions as well as those containing further details about the simulations are freely available upon request to the corresponding author (TTG). \label{foo:code}} Our combined $N$-body/HD simulation includes the effect of radiative cooling of the gas by hydrogen, helium and heavy elements, and heating by the cosmic ultra-violet radiation background, but it does not include additional sub-grid physics such as star formation or feedback of any kind. This implies that our model as is {\em cannot} be used to make any statements about the details of the ionisation state of the gas, or the details (formation, age, chemical composition and its distribution) of the stellar components of the galaxies.

We kindly refer the reader to our previous work \citep[][and references therein]{tep19a} for more details on our simulation technique.

The orbital history that results from our $N$-body simulation is displayed in the right panel of Fig.~\ref{fig:orbit} and is indicated by the plus signs (+), along with the corresponding results from our forward semi-analytic integration (solid curve). The values of the corresponding 3D position and the 3D velocity are given in Tab.~\ref{tab:orbit}. The relative position and the relative velocity of the galaxies in our $N$-body simulation are calculated by estimating the centre of mass and centre of velocity of each galaxy's stellar disc at each time step.\footnote{Using the position of centre of mass of their respective DM halo yields virtually identical results.} We do this in an iterative way until their respective value convergence within some tolerance to take into account strong anisotropies in both the mass and the velocity distribution caused by tidal distortion. The calculation of the centre of velocity is performed in the same way. However, it bears the additional complication that the velocity of the individual stars is superimposed on the barycentre's motion as they revolve around the potential's centre. This explains the larger departure of the relative speed relative to the SAOI result. In contrast, the relative position in the $N$-body model faithfully follows the SAOI result.

The results of Fig.~\ref{fig:orbit} and Tab.~\ref{tab:orbit} demonstrate that orbital history obtained from our $N$-body simulation are consistent with the SAOI result. This is indeed remarkable considering all the simplifying assumptions factored into our SAOI calculations. It must be mentioned though that the difference between the present-day orbital parameters obtained from our $N$-body simulation and their observed counterpart is larger than the difference between the latter and our SAOI result. Nonetheless, we are confident that the physical processes at play relevant to the dynamical evolution of the M31/M33 system (gravity, dynamical friction, tidal stripping) are captured correctly in our $N$-body model. We believe that extensive tuning of the free parameters of the model (galaxy masses; their spin orientation, of each galaxy; $\epsilon$, etc.) -- constrained by observations where possible -- may yield a better match between the present-day orbital parameters of the M31/M33 system as observed and their corresponding values resulting from an $N$-body simulation. That said, a more thorough exploration of the orbital parameter space is left for future work.

We thus consider the agreement between the observed present-day orbital parameters of the M31/M33 system and the result from our $N$-body model good enough for our present purpose, and proceed to model the infall of M33 into M31 using a combined $N$-body/HD simulation.

%--------------------------------------------------------------------------------------------------------------------------------------------------------------------------------
% FIGURE: Stellar morphologies
\begin{figure*}
\centering
\includegraphics[width=0.33\textwidth]{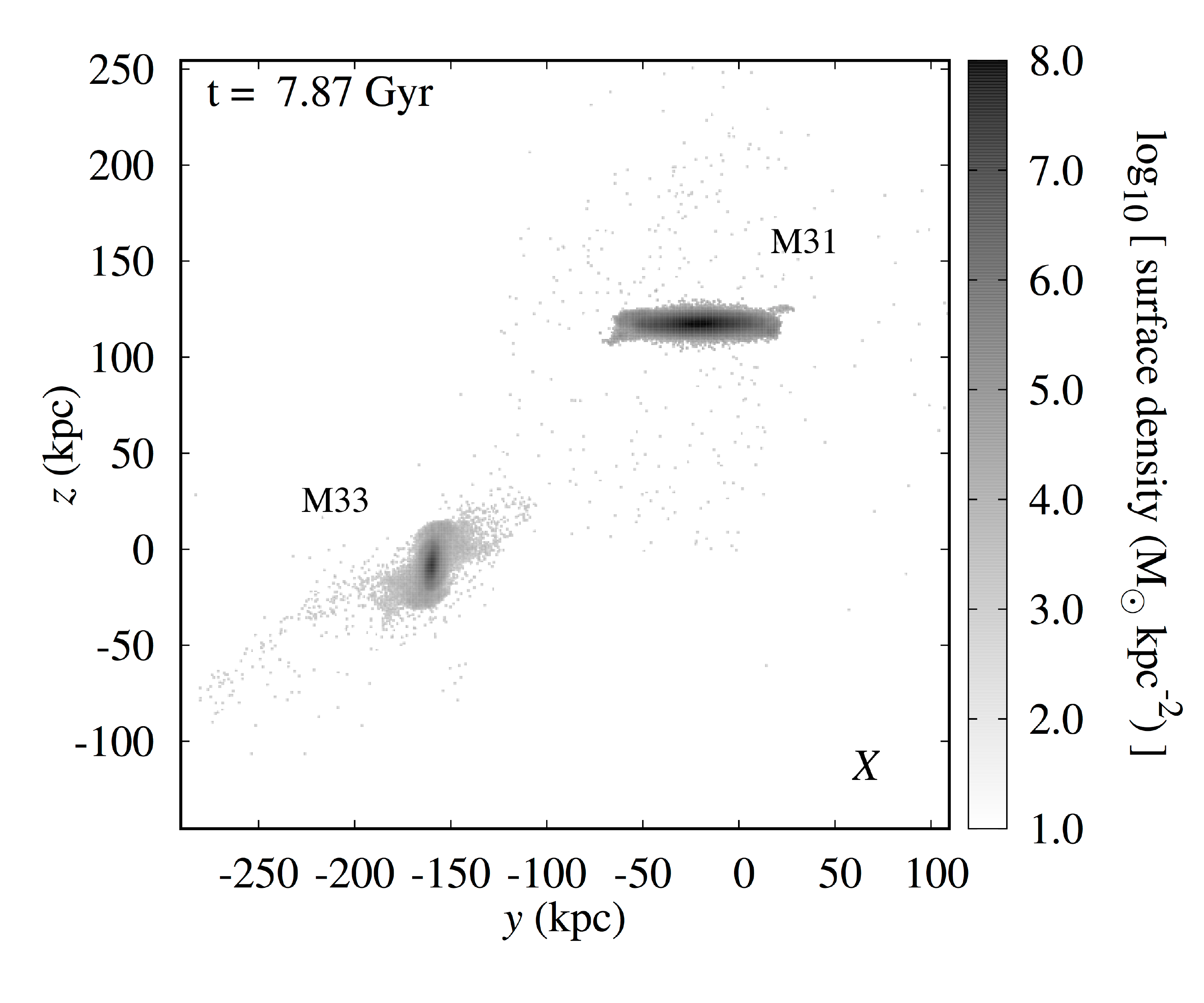}\hfill
\includegraphics[width=0.33\textwidth]{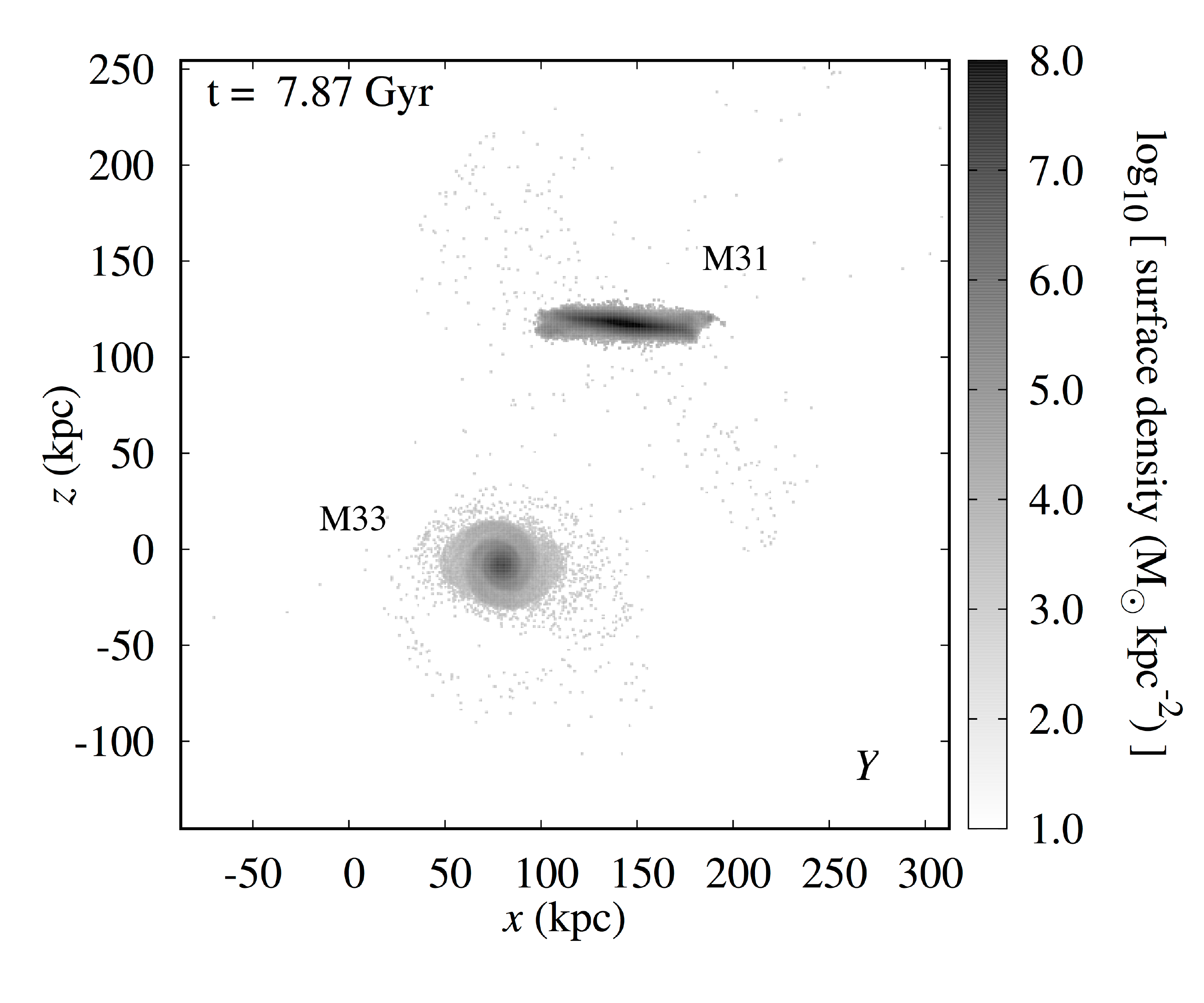}\hfill
\includegraphics[width=0.33\textwidth]{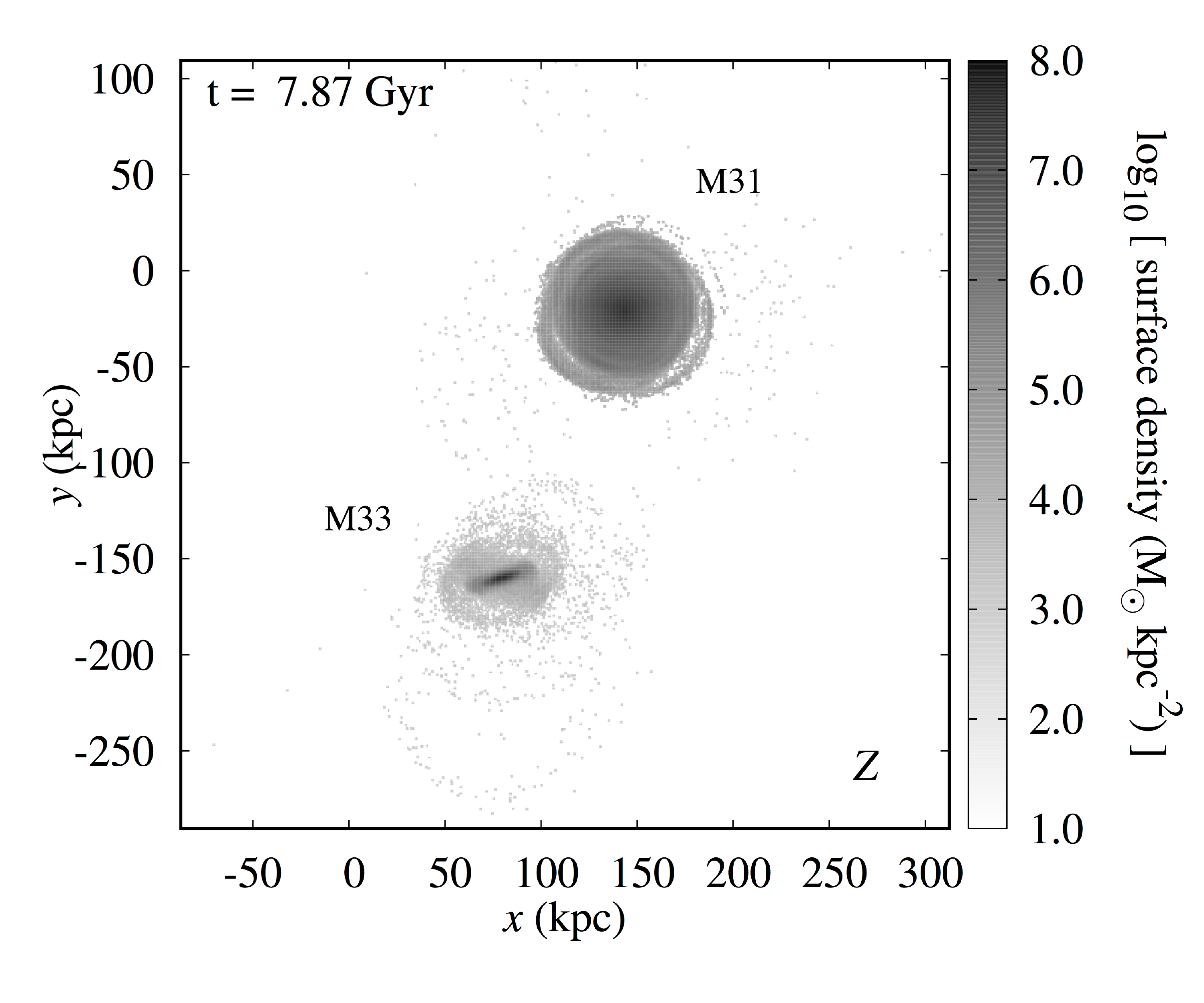}
\vspace{-10pt}
\caption[  ]{ Distribution of stars in the M31/M33 system at the present epoch, i.e., $\sim 7$ Gyr after M33 first crossed M31's virial radius and $\sim 6.5$ Gyr since their closest encounter, in our $N$-body/HD simulation. From left to right, each column corresponds to the full 3D state of the system projected along the $x$-, $y$-, and $z$-axis, respectively. The gray scale indicates the logarithmic value of the stellar surface density in units of $\Msun ~\kpc^{-2}$. Note that the coordinate frame has its origin at the initial (i.e., at the start of the simulation) position of M31's centre of mass. }
\label{fig:stars0}
\end{figure*}
%--------------------------------------------------------------------------------------------------------------------------------------------------------------------------------

%--------------------------------------------------------------------------------------------------------------------------------------------------------------------------------
\subsection{A full gas-dynamical model} \label{sec:hd}

In our $N$-body/HD model, each of the galaxies is assumed to host a gas disc, in addition to the collisionless components described in the previous section. \citet[][]{sem18a} demonstrated the importance of a hot halo around M31 in order properly account for the effect of ram pressure onto the gas around M33. Therefore, our M31 model includes a hot halo as well. We note that \citet[][]{sem18a} did not specify the metallicity of the hot halo (M31) or the gas disc (M33). This is important if the effect of radiative cooling of the gas is accounted for, as we do. Guided by the corresponding values of the Milky Way \citep[][]{mil15a,hou00a}, we adopt a value of 0.3 solar for the metallicity of the M31's gas components, and a slightly lower value of 0.2 solar for M33's gas disc. The galaxies are embedded in a diffuse background medium\footnote{The density background is likely much more diffuse than the intergalactic medium (IGM) in the Local Group. We opt for such a tenuous medium to deter the gas filling the simulation volume from collapsing over the course of the simulation.} with a mean temperature $\sim 10^6$ K and a density $\sim 10^{-7}$ \pcc.\\

Rendered animations of the evolution of the M31/M33 system can be found at \url{http://www.physics.usyd.edu.au/\~tepper/proj\_m31m33.html}.

%--------------------------------------------------------------------------------------------------------------------------------------------------------------------------------
\section{Tidal structures: Echoes of a past interaction} \label{sec:tidal}

The stellar halo of M31 is observed to feature a rich variety of structures which are likely the result of an extended period of interaction with -- and accretion of -- smaller systems. The most remarkable of these structures, the Giant Stellar Stream \citep[GSS;][]{iba01a}, extends well beyond a projected distance of 40 kpc from M31's central region along the line towards M33. Using $N$-body models, \citet[][]{far06a} have attributed the origin of this structure to the recent interaction  ($\sim 750$ Myr ago) of M31 with a relative small companion ($\sim 10^9$ \Msun). Others \citep[e.g.][]{ham18a} argue for a major merger event as the source of the GSS. In view of the new orbital history presented here, it is worth exploring whether the origin of this structure could be attributed to a close encounter with M33 in the distant past

Similarly, the observed distribution of gas around M31 strongly suggests that this galaxy has had a number of interactions with its companions. The \HI\ filament along the projected line linking M31 with M33 \citep[][]{bra04a,loc12a,ker16a} is of particular interest. This gaseous structure has been interpreted as either resulting from an interaction between these two galaxies \citep[e.g.][]{bek08a} or the product of condensation within an intergalactic filament \citep[][]{wol13a,wol16a}. The latter interpretation has gained support lately as a result of the claims that M33 may be on its first approach to M31 \citep[e.g.][]{van19a}. But our orbital calculations anew render the interaction scenario a plausible alternative as the potential origin of the \HI\ filament, as we discuss below.

%--------------------------------------------------------------------------------------------------------------------------------------------------------------------------------
\subsection{Stellar morphologies} \label{sec:star}

The present-day distribution of stars in the simulated M31/M33 system is displayed in Fig.~\ref{fig:stars0}, projected along three orthogonal directions. This choice has no other motivation than simplicity. While none of the three resulting views fully matches the configuration of M31/M33 as seen on the sky, this is of no concern as we still have not exploited the freedom given by the isotropy of the simulation volume, which allows is to rigidly rotate the $xyz$-coordinate system in whichever way we see fit, nor have we adjusted the relative orientation of the initial spin of each galaxy with respect to their infall orbital parameters. With this in mind, we proceed with a qualitative comparison between our simulation results and observations.

%--------------------------------------------------------------------------------------------------------------------------------------------------------------------------------
% FIGURE: Gas streams
\begin{figure*}
\centering
\includegraphics[width=0.33\textwidth]{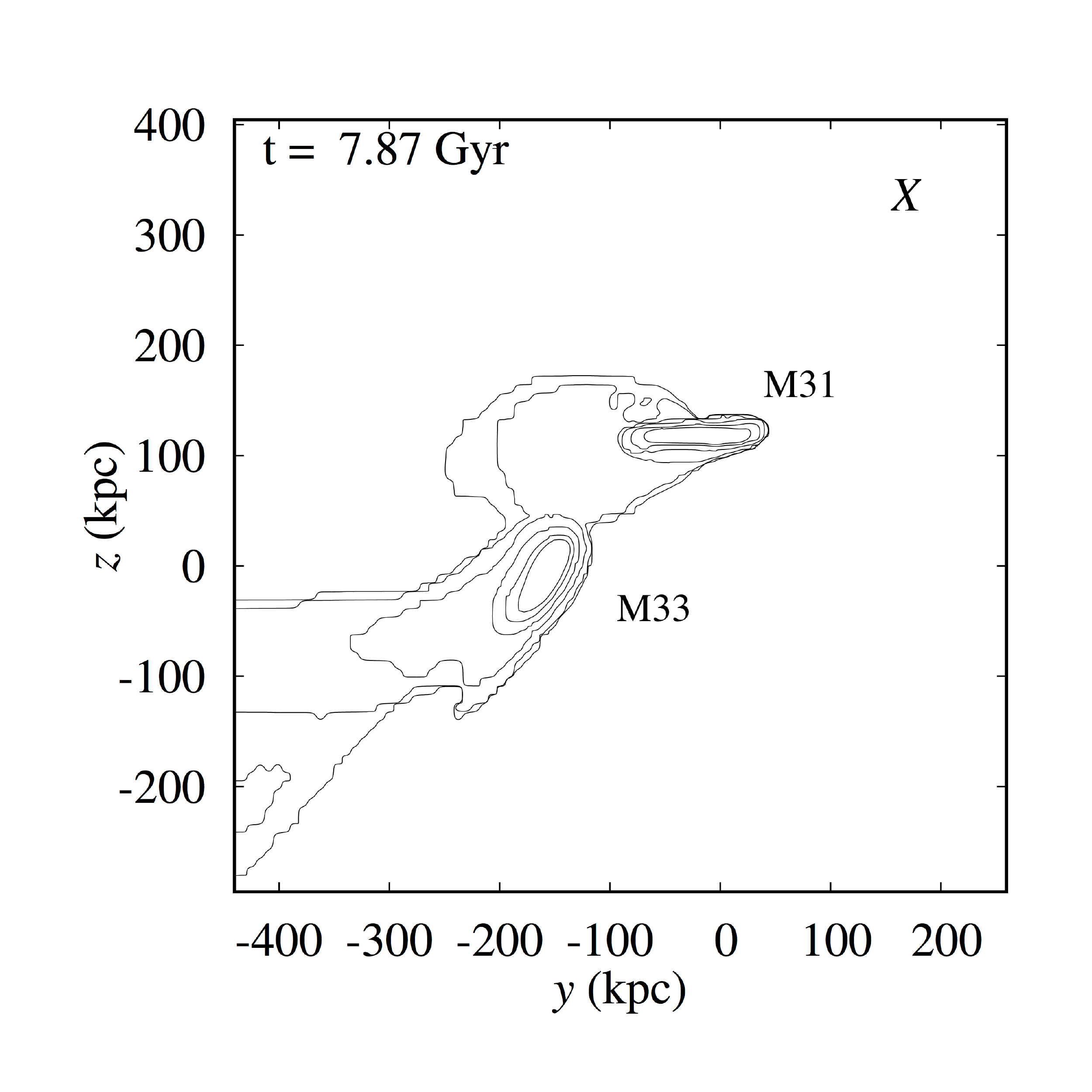}\hfill
\includegraphics[width=0.33\textwidth]{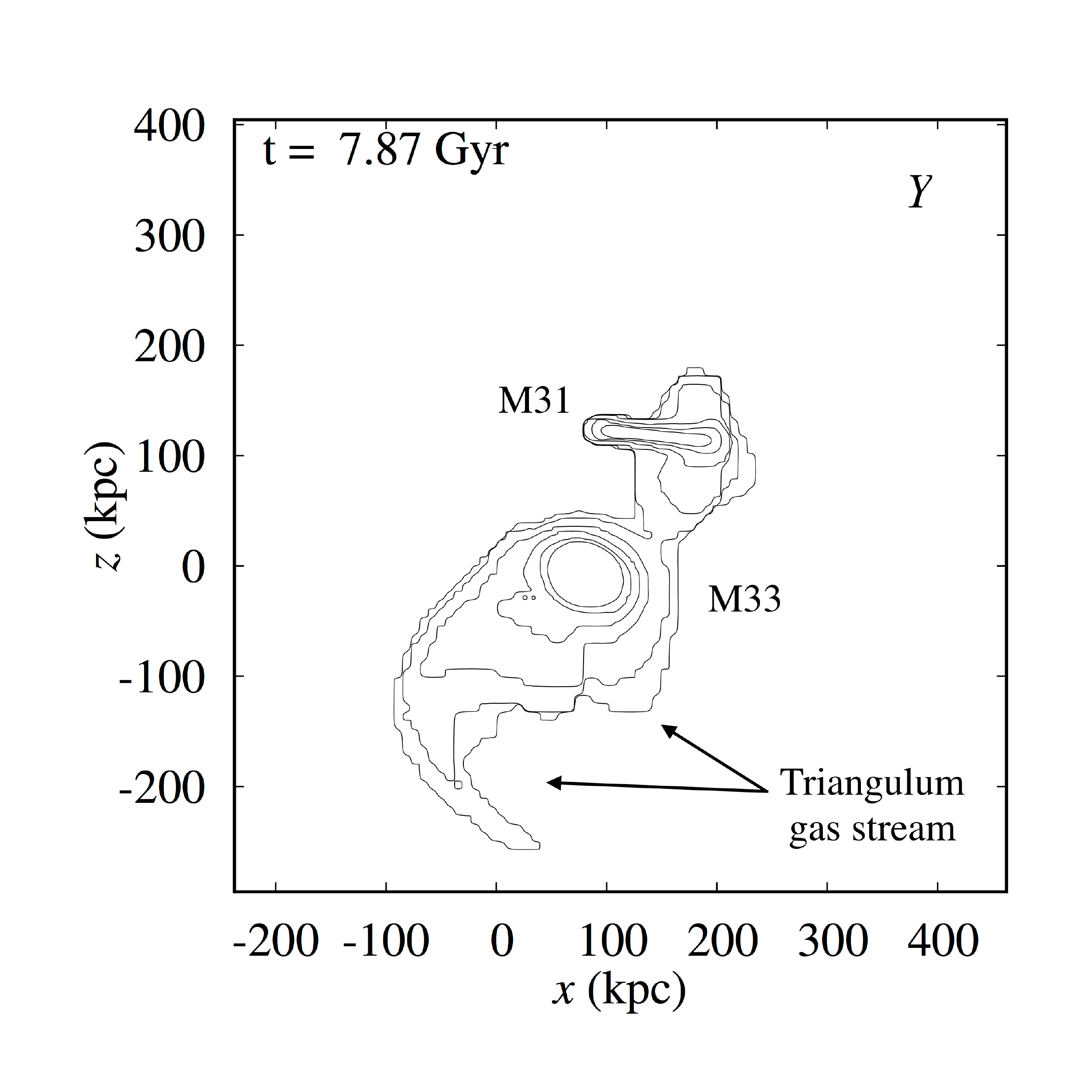}\hfill
\includegraphics[width=0.33\textwidth]{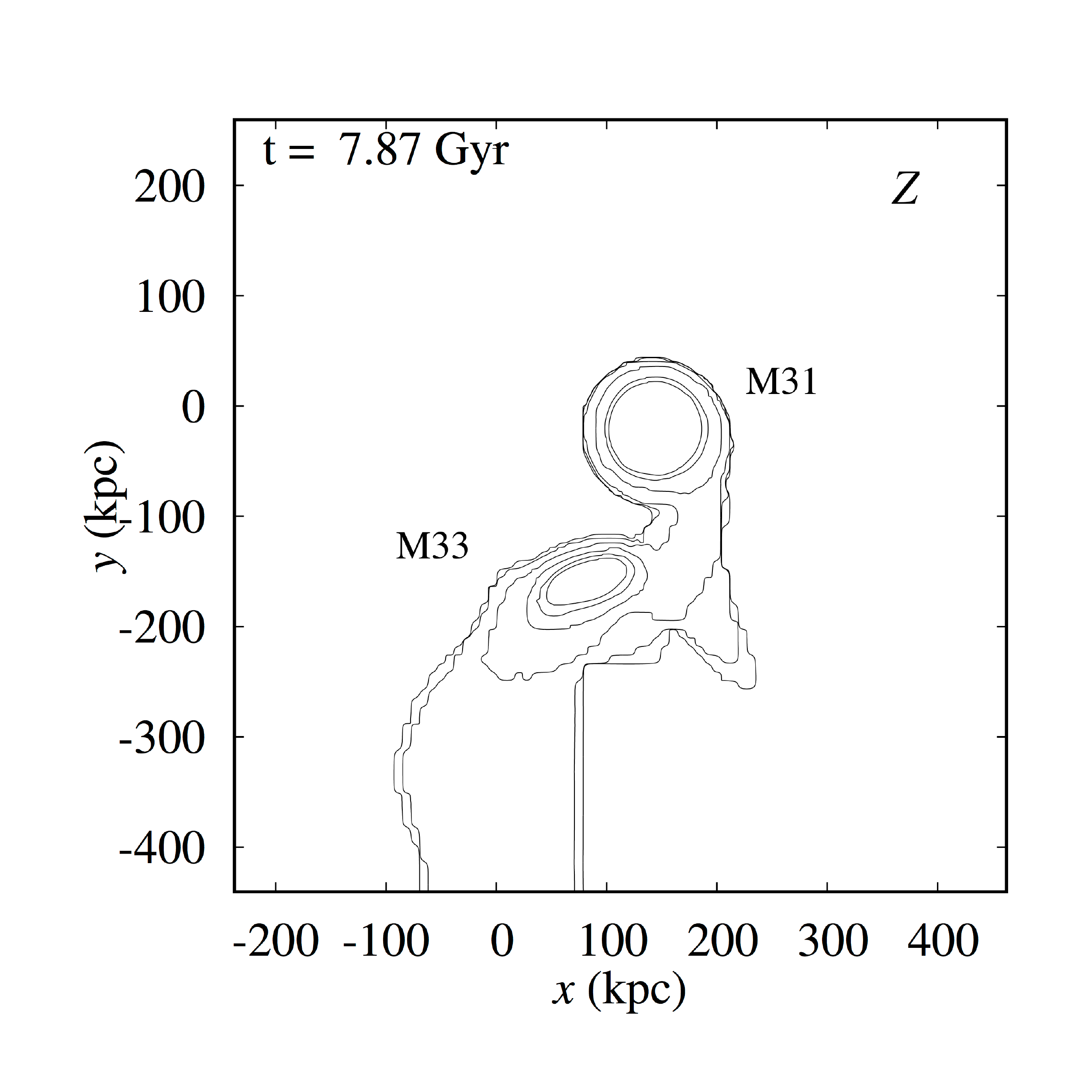}
\vspace{-10pt}
\caption[  ]{ Distribution of gas in the M31/M33 system at the present epoch, i.e. $\sim 7$ Gyr after M33 first crossed M31's virial radius and $\sim 6.5$ Gyr since their closest encounter, in our $N$-body/HD simulation. From left to right, each column displays the density of gas projected along the $x$-, $y$-, and $z$-axis, respectively. Only the gas initially bound to M31 and M33 is shown. The contour levels indicate, starting from the innermost, a logarithmic total gas column density ($\log [ \Ng / \psc ]$) from 19 dex to 17 dex in steps of 0.4 dex. Note the difference in physical scale with respect to the panels in Fig.~\ref{fig:stars0}. Gas structures reminiscent of the observed \HI\ cloud complex north of M31 and of the M31-M33 \HI\ filament are apparent in the central panel, in particular. The prominent, diffuse gas structure extending from M33 {\em away} from M31, here referred to as the `Triangulum stream', is yet to be observed. The stream has been tagged in the central panel only but it is clearly visible in the other projections as well. Note that the coordinate frame has its origin at the initial (i.e. at the start of the simulation) position of M31's centre of mass. }
\label{fig:gas0}
\end{figure*}
%--------------------------------------------------------------------------------------------------------------------------------------------------------------------------------

The most apparent characteristics of the simulated M31/M33 systems visible in the top row of Fig.~\ref{fig:stars0} are the lack of notorious stellar substructure around M31. In contrast, M33 displays a number of tidal stellar features (see Sec.~\ref{sec:m33}). The close encounter between these galaxies in our model does result in both systems displaying strong tidal features around each other shortly thereafter. But M31 settles quickly so that it appears fairly undisturbed today, in contrast to observations. In particular, the simulated M31 does not display a structure reminiscent of the Giant Stellar Stream. Since we have ignored the presence of M31's satellites other than M33, the disagreement between our model and the data is consistent with the belief that most of the stellar substructure around M31 observed today is either due to interactions with its low-mass companions \citep[e.g.][]{far06a,wil15a} or the result of an ancient major merger \citep[][]{ham18a}.

%--------------------------------------------------------------------------------------------------------------------------------------------------------------------------------
\subsection{Gas streams} \label{sec:gas}

The simulated M31/M33 system displays a spectacular gas distribution at the present epoch (Fig.~\ref{fig:gas0}). Both galaxies show tidal gas streams around them, extending for at least tens of kpc from their respective centre. Yet, the gas distribution around M31 appears more settled compared to M33, as does its stellar component (see previous section). The prominent gas stream extending from M31 {\em away} from M33 is reminiscent of the \HI\ stream detected north of M31 (see \citealt[][]{bra04a}, their figure 9; \citealt[][]{wol16a}, their figure 1).

A panoramic view of the gas around the simulated M31/M33 system is shown in Fig.~\ref{fig:gas0}. M31 features a second prominent stream which extends all the way to M33, akin to the observed M31-M33 \HI\ filament, without a stellar counterpart. We note that the formation epoch of this gas bridge dates back to the epoch of the close encounter between the galaxies, and apparently survived up to the present-day in our simulation.

In addition to the gas bridge between the galaxies, M33 features a gas stream that trails from the galaxy for more tens of kpc -- henceforth referred to as the `Triangulum stream'. This gas stream is not apparent in any of the \HI\ maps available to date, although it may be discernible in the data presented by \citet[][]{put09a} and \citet[][]{kee16a}. Our model suggests that the Triangulum stream is very diffuse, with {\em total} gas column densities $\Ng \lesssim 10^{18}$ \psc. Thus, we speculate that, if real, it has not been observed in \HI\ emission because of its low \HI\ column density, which may result from an overall low gas column density and/or a high ionisation fraction of the gas; or because its distribution could be clumpy \citep[][]{bla17a}. However, without an appropriate treatment of the ionisation state of the gas accounting for its chemical composition, for the ionisation field of the M33 and for the ultra-violet background, and a sufficiently high grid resolution, we cannot as yet make a more quantitative statement about the \HI\ content of the Triangulum stream. Such an endeavour extends beyond the scope of the present paper and is left for future work.

The reason for the more extended distribution of gas compared to the stars, and the existence of gas streams with no stellar counterpart is that, in addition to tidal forces, the gas is subject to hydrodynamic forces exerted by the IGM as the galaxies move along their orbit. Therefore, the properties of the gas streams will largely depend on the properties of the IGM. In particular, a IGM significantly denser than we have adopted in our model will likely yield to the formation of denser trailing tidal gas streams, and weaker leading ones around M33 as a result of the enhanced ram pressure ahead of the galaxy \citep[q.v.][]{tep19a}.

The initial extension of the gas disc relative to the stellar disc may also play a role in the different configuration of gas streams relative to stellar streams \citep[][]{mih01a}. In our model we have assumed indeed very extended gas discs ($\gtrsim 40$ kpc) for both galaxies. In the case of M31, this does not seem implausible, given that the Milky Way is believed to have a gas disc that extends out to $\sim 60$ kpc \citep[][]{kal09a}. The \HI\ disc of M33 is observed to extend at least out to $\sim 25$ kpc \citep[][]{cor14b}, and it is not unreasonable to assume that its ionised edge may extend beyond that \citep[][]{bla17a}. It is also possible that its extension at infall was much larger than that. Thus, while most of the stars sit close to the potential's centre, a non-negligible fraction of the gas is loosely bound and gets more easily stripped.\\

To sum up, our model indicates that the Giant Stellar Stream is unlikely the result of an encounter in the distant past between M31 and M33. But the \HI\ filament observed between these galaxies can be naturally explained in this scenario, at least qualitatively. Our model suggests that this structure is ancient, dating back to the epoch of the close encounter between the galaxies, some 6.5 Gyr ago. Incidentally, our model predicts the presence of a yet to be observed gas stream, the Triangulum stream, that trails M33 as it moves towards M31.

%--------------------------------------------------------------------------------------------------------------------------------------------------------------------------------
\subsection{The morphology of M33} \label{sec:m33}

%--------------------------------------------------------------------------------------------------------------------------------------------------------------------------------
% FIGURE: M33's stellar (and gas) structures: $N$-body / $N$-body+HD
\begin{figure*}
\centering
\includegraphics[width=0.33\textwidth]{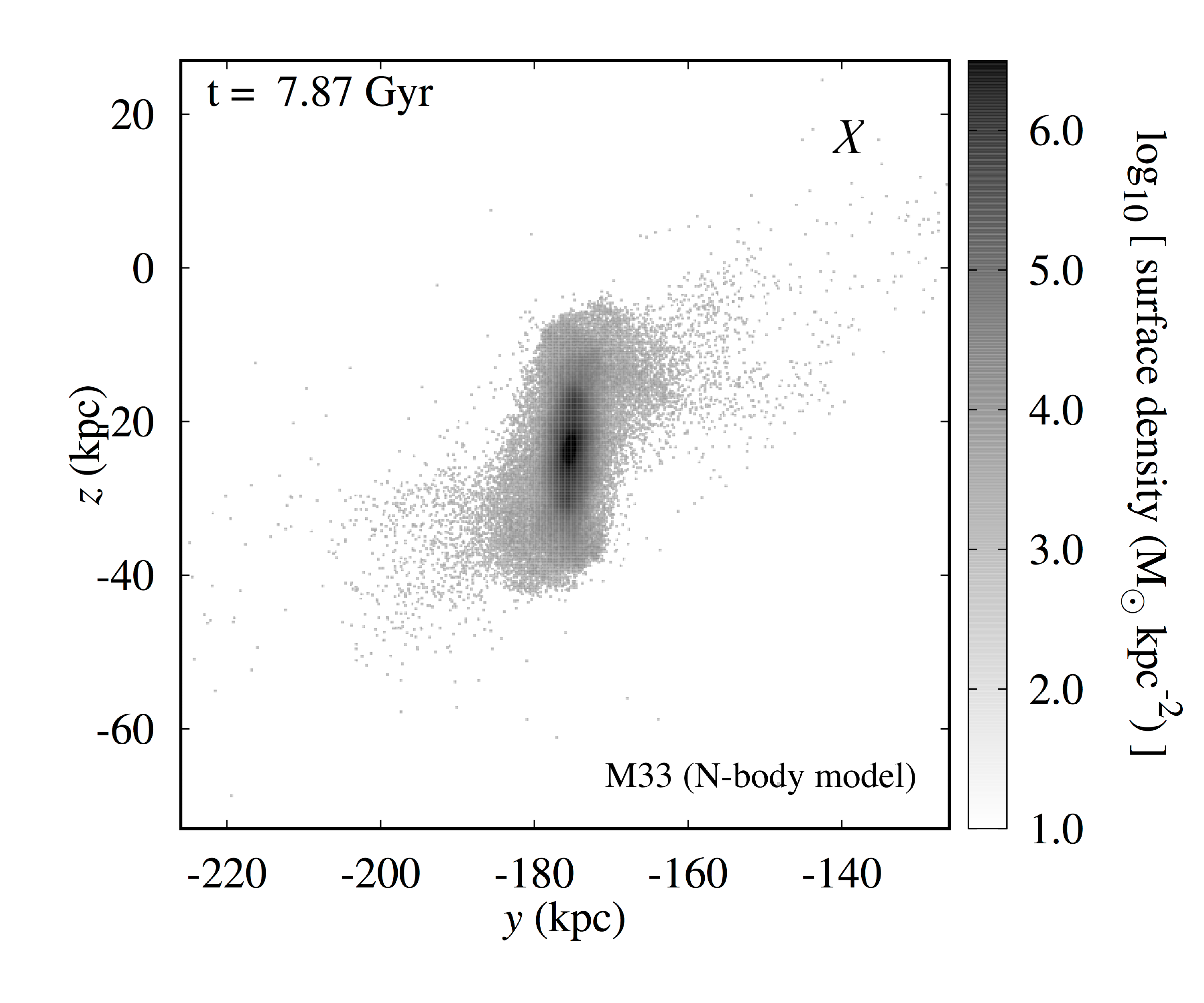}\hfill
\includegraphics[width=0.33\textwidth]{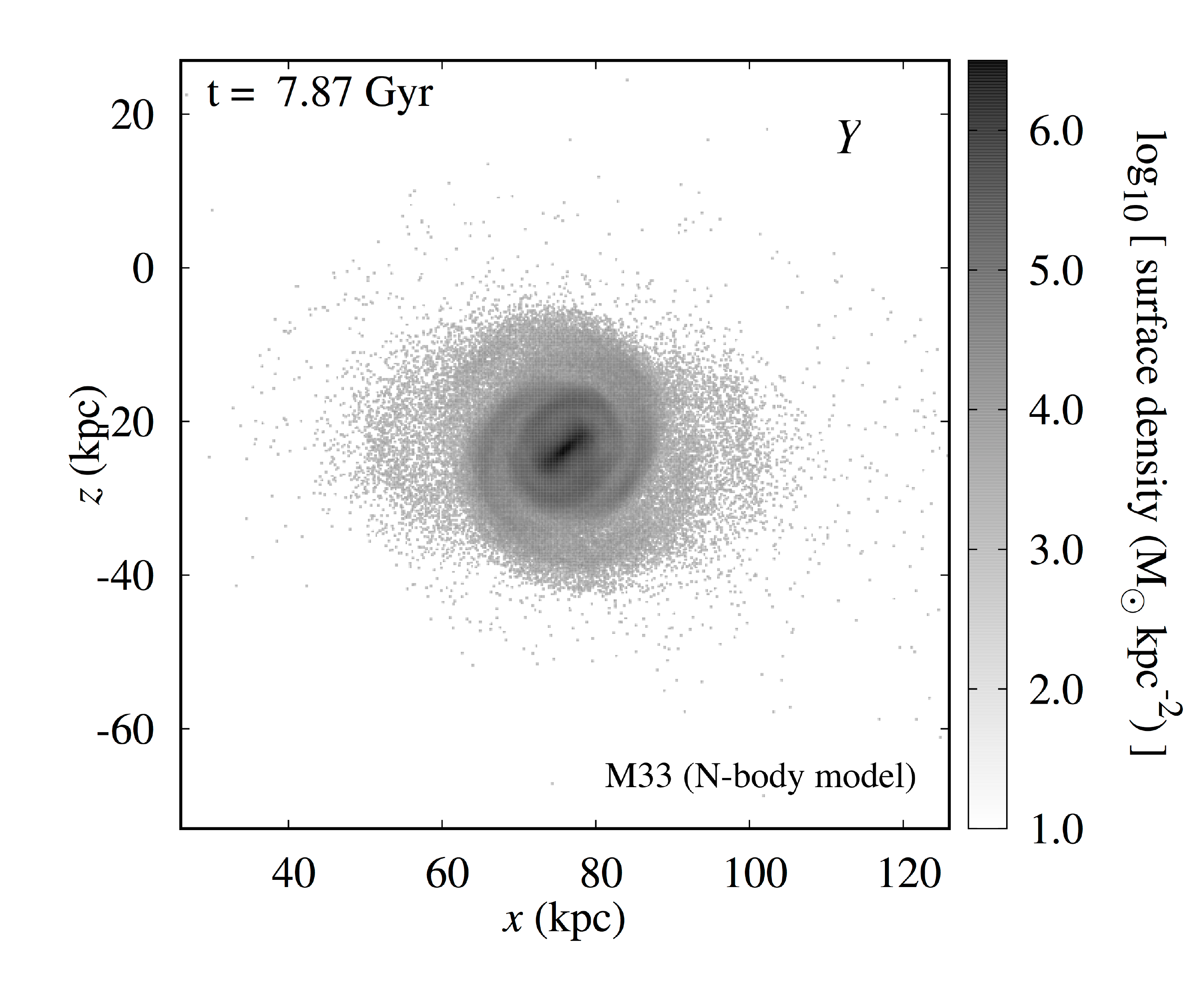}\hfill
\includegraphics[width=0.33\textwidth]{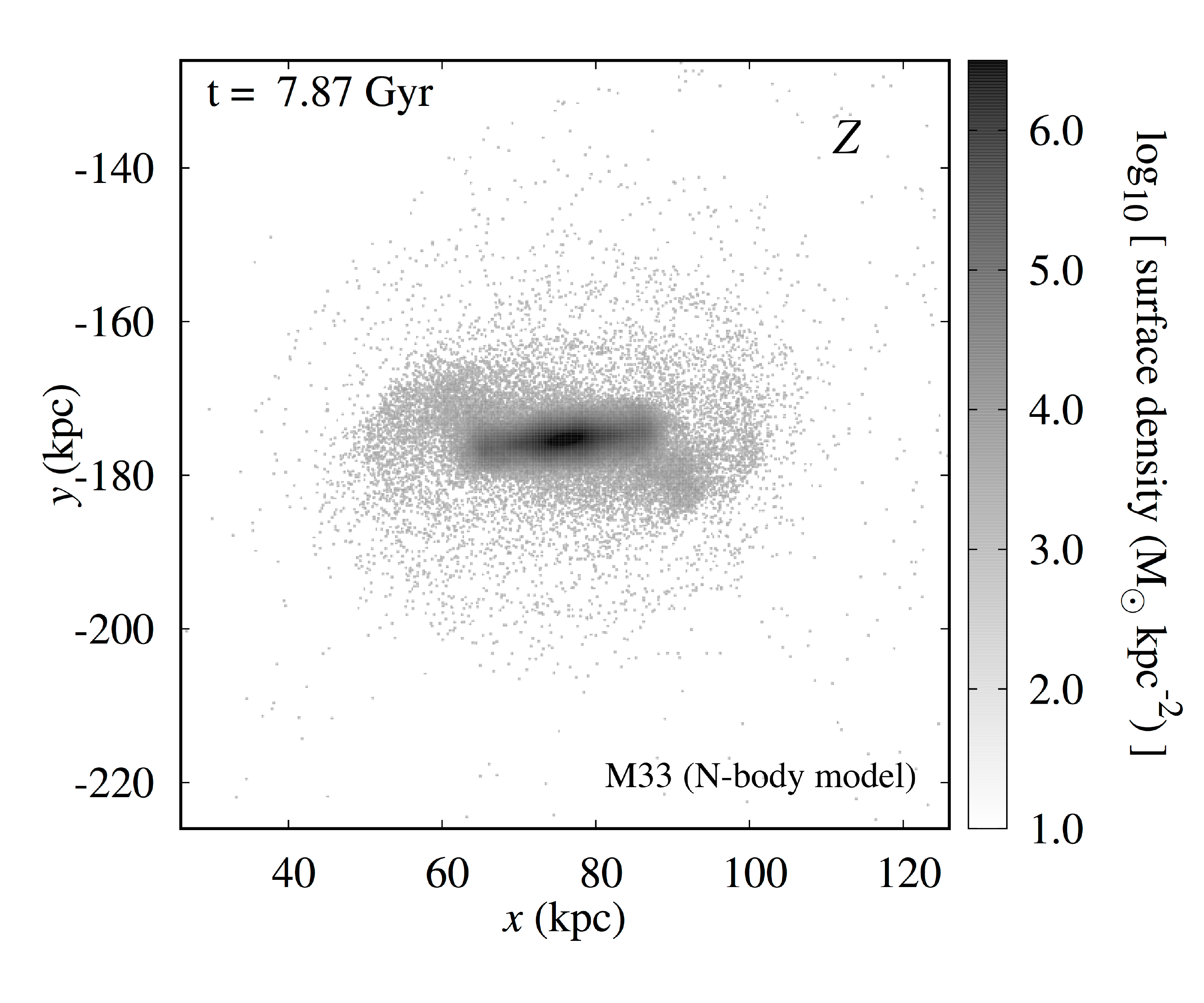}
\includegraphics[width=0.33\textwidth]{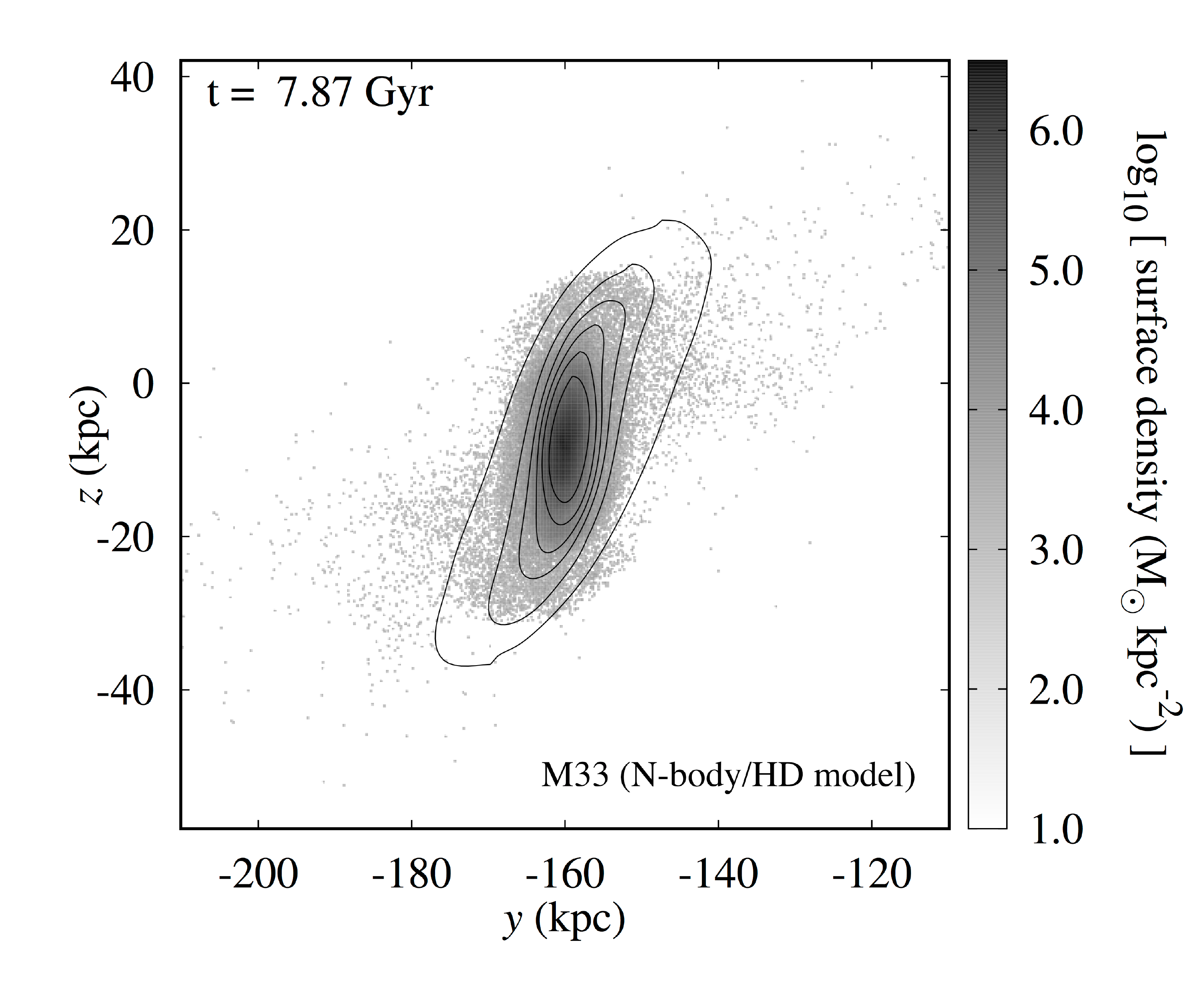}\hfill
\includegraphics[width=0.33\textwidth]{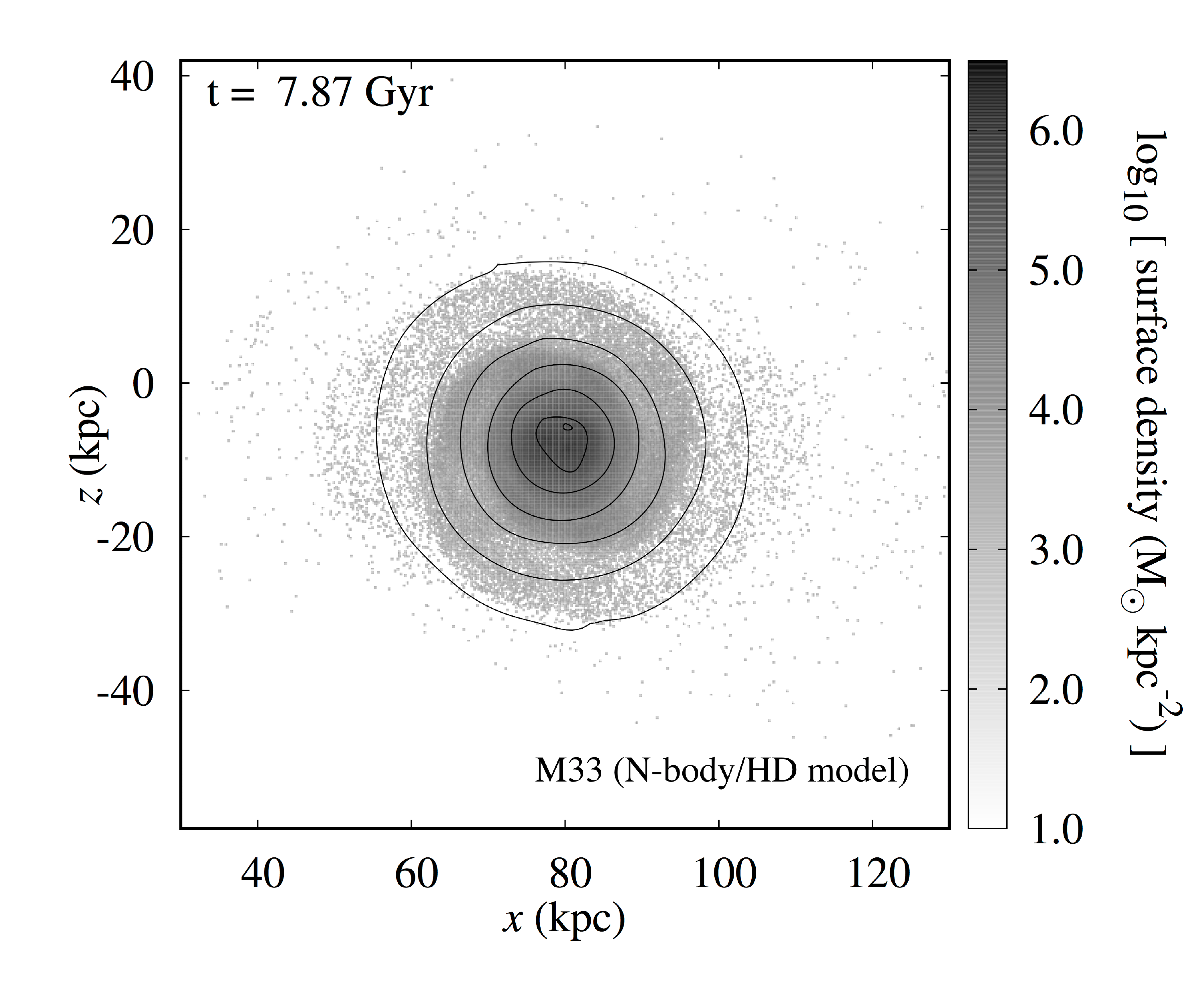}\hfill
\includegraphics[width=0.33\textwidth]{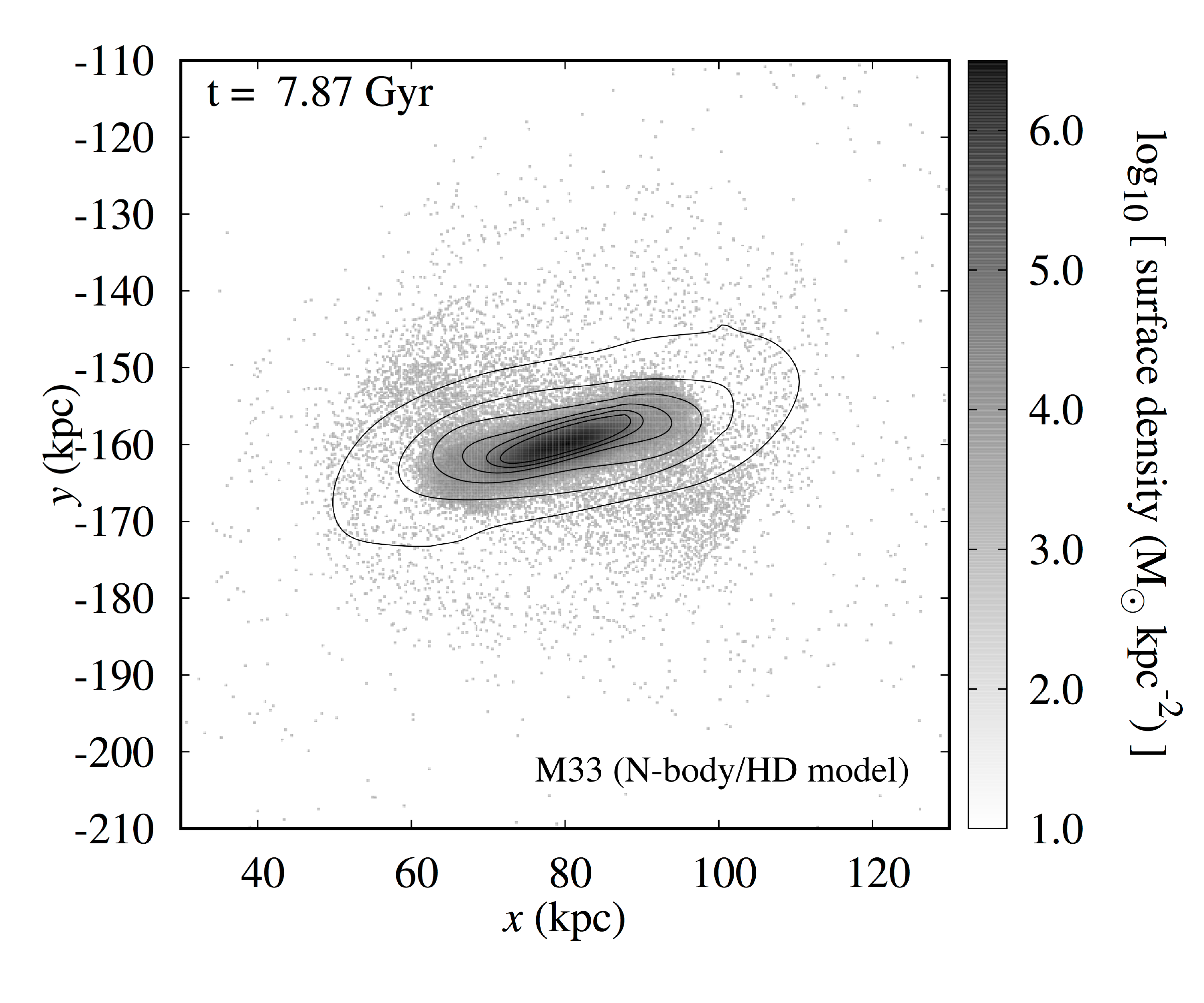}
\caption[  ]{ Simulated distribution of stars (and gas) in M33 at the present epoch, i.e. $\sim 7$ Gyr after M33 first crossed M31's virial radius and $\sim 6.5$ Gyr since their closest encounter. From left to right, each panel displays the configuration of M33 projected along the $x$-, $y$-, and $z$-axis, respectively. Top: $N$-body simulation. The gray scale indicates the logarithmic value of the stellar surface density in units of $\Msun ~\kpc^{-2}$. Note the warp of the disc and the presence of a bar at the galaxy's centre. Bottom: Combined $N$-body/HD simulation. The gray scale has the same meaning as in the top panels. The contours indicate the total gas column density; each of the levels corresponds, starting from the innermost, to a logarithmic total gas column density ($\log [ \Ng / \psc ]$) from 21.4 dex to 9.4 dex in steps of 0.4 dex. The warp of the disc is apparent, but there is no bar discernible. Note that the coordinate frame has its origin at the initial position of M31's centre of mass.}
\label{fig:m33gasstars0}
\end{figure*}
%--------------------------------------------------------------------------------------------------------------------------------------------------------------------------------

Although somewhat beyond the scope of our present study, we would like to highlight some interesting features in the morphology of the simulated M33. First, it is worth noting the difference in the stellar morphology between the pure $N$-body model and the $N$-body/HD model. In the former, M33 quickly forms a bar shortly after its close encounter with M31 which survives up to the present day in the simulation (Fig.~\ref{fig:m33gasstars0}, top row). In contrast, the $N$-body/HD model M33 does not (Fig.~\ref{fig:m33gasstars0}, bottom row).

The latter is relevant in the context of the M33's present-day dynamical state, which has recently been revisited by \citet[][]{sel19a}. Performing an extensive study using $N$-body/HD simulations, they have found that any dynamical model broadly consistent with M33's observed properties is unstable to the formation of a strong bar within a short time-scale ($< 1$ Gyr), contrary to what is observed. We note that their models consider M33 only in isolation. Clearly, our result is in tension with their findings. A preliminary analysis of the structural properties of both our $N$-body and $N$-body/HD M33 models performed in cooperation with J.~A.~Sellwood has revealed that the stellar disc in either model {\em in isolation} is stable against bar formation. This stems from two factors: 1) The radial profile of Toomre's stability parameter $Q$, which is $\approx 2$ in the inner ($R \lesssim 4$ kpc) disc, but rises quickly to $\gtrsim 5$ by $R \approx 8$ kpc; and 2) the value of its scaleheight, which is roughly a factor 1/5 of its scalelength; i.e., the disc is relatively warm and thick. These findings are in agreement with \citeauthor[][]{sel19a}'s conclusions.

The fact that our $N$-body model {\em does} develop a bar after the close encounter with M31 is consistent with the idea that the instability in an otherwise stable disc can be triggered by the tidal interaction with another galaxy \citep[e.g.][]{lok14a}. The fact that the $N$-body/HD model {\em does not} feature a bar at the present epoch demonstrates that, in the event of a tidal interaction, the presence of gas may significantly alter the dynamical properties of the disc. This may happen either because: 1) the gas increases the effective value of $Q$ at all radii \citep[][]{rom13a}, thus adding stability to the disc and preventing the formation of a bar from the outset; or 2)  because gas flows towards the disc's centre after the interaction, thereby increasing the central mass concentration, and eventually leading to the dissolution of an existing bar \citep[][]{pfe90a}.

The simulated M33 in our $N$-body/HD run displays a warp both in its stellar disc and in its gas disc (Fig.~\ref{fig:m33gasstars0}, bottom row), in qualitative agreement with what is observed. It is interesting however that while the warp of the gas disc and the stellar tidal streams roughly align with each other along one projection (left panel), they appear anti-aligned along a different one (right panel). These configuration is likely particular to our initial orbital parameters and the spin orientation of M33 relative to its orbit. Nonetheless, these results highlight the complexity of the structure that results even in the simplest of models.

\citet[][]{van19a} argue, reasonably, that in lack if a close encounter with M31, the stellar and gaseous warps and tails of M33 cannot be the result of tidal forces exerted by M31. This in turn poses a puzzle on the origin of M33's disturbed morphology. They discuss a number of alternative scenarios, among them the possibility that M33's disturbed structure may have arisen from the interaction with a population of yet-to-be-detected low-mass companions \citep[see][]{pat17b,pat18a}. Here we have demonstrated, as have others before us \citep[e.g.][]{mcc09b,sem18a}, that these structures can naturally be explained within an interaction scenario.

%--------------------------------------------------------------------------------------------------------------------------------------------------------------------------------
\section{Summary \& final remarks} \label{sec:sum}

We have shown that an orbital history featuring a close encounter between M31 and M33 constrained by the central values of the \gaia\ DR2 PM measurements exists, provided the effect of dynamical friction and dynamical mass loss are self-consistently accounted for. In addition to highlighting the importance of tidal stripping, our orbital calculation shows that assumptions generally adopted about the value of poorly constrained parameters (e.g., $\epsilon$) which enter the semi-analytic calculation of the orbit do not necessarily hold. In this respect, we stress the importance of $N$-body simulations to validate the results obtained from the application of pure semi-analytic methods. This may imply the need to revisit previous orbital analyses of galaxy pairs which have ignored dynamical mass loss and/or the use of $N$-body simulations.

In view of the somewhat laborious steps involved in our orbit calculation, we have refrained from performing an exploration of the orbital parameter space allowed by the \gaia\ DR2 PM and their associated uncertainty. Without such an analysis it is not possible to make a statement about the statistical significance of our orbit. Earlier studies \citep[][]{pat17b} have shown that a light ($< 10^{11}$ \Msun) M33 disfavours orbits featuring a close encounter with M31. We have not explored the mass ranges they have for either M31 or M33; rather, here we have adopted the intermediate (in terms of mass) of their M33 models, and the heaviest M31 model they considered. Based on \citet[][]{pat17b}'s extensive orbital analysis, \citet[][]{van19a} concluded that the \gaia\ DR2 PM exclusively support orbital histories where M33 is on its first approach to M31. The orbit presented here is at odds with this claim. Therefore, and despite the lack of a statistical analysis of the allowed orbits, we speculate that orbital calculations which adopt a heavier M33 or use a different set of PM measurements \citep[e.g. the weighted \gaia\ / {\em HST} PM measurements; cf.][]{van19a} would likely yield many more orbital solutions that allow for a close encounter between these galaxies.

Our present study and previous similar studies together demonstrate that the two diametrically different orbit families, i.e., a first-infall and one featuring a close encounter in the distant past (`second infall') consistent with the \gaia\ DR2 PM are both in fact possible, and further investigation beyond an orbital analysis is required to discriminate between these scenarios. For instance, our calculations show that a second-infall solution as the one presented here requires M33 to have lost roughly half its mass to tidal stripping. In contrast, a first-infall solution is consistent with an insignificant mass loss. Therefore, inferring M33's mass evolution over the last $\sim 8$ Gyr should prove a useful discriminant, albeit a hard one to infer. The study of its present-day satellite population -- provided it existed at any epoch -- could be used as a proxy for the analysis of its own mass evolution as a way of constraining its orbital history \citep[][]{pat18a}.

Despite the possibility of both orbit families, we find a wealth of evidence firmly supporting the belief that M33 had a strong tidal interaction with M31 in the past. This includes the disturbed stellar morphology of M33 and its warped disc, its star formation history (SFH), and the \HI\ filament observed between these two galaxies. Indeed, with no obvious perturber other than M31, it is difficult to understand how M33's disturbed structure came about to be \citep[but see][]{pat18a}. Similarly, the idea that the M31-M33 \HI\ filament may result from the condensation of IGM gas, while not unreasonable, may require very special conditions \citep[e.g.][]{bin09a}. In contrast, a close encounter between these galaxies may naturally explain these observations -- at least qualitatively, as others have argued before us (albeit using different orbital constraints, e.g., \citealt{mcc09b}; \citealt{sem18a}, and less sophisticated models, e.g., \citealt{bek08a}).

Using infrared data, \citet[][]{jav17a} found evidence for two epochs during which M33's star formation rate was enhanced by a factor of a few, one of which started $\sim$6 Gyr ago and lasted for $\sim$3 Gyr, producing more than 70 per cent of the total mass in stars. The earliest event agrees with the age of the dominant stellar population in the central region of M33 determined from optical data \citep[][]{wil09b}. The epoch of the pericentric passage of M33 around M31 suggested by our orbital solution ($\sim 6.5$ Gyr ago) corresponds to a redshift $z \approx 0.75$ in a 737-cosmology, i.e., $(h, \Omega_b, \Omega_\Lambda) = (0.7, 0.3, 0.7)$. This event just precedes the earliest episode of enhanced star formation in M33, suggesting it was triggered by the close encounter between these galaxies \citep[e.g.][]{may01a}.

%Thus, an M31/M33 interaction scenario consistent with the \gaia\ DR2 PM provides a natural and unifying picture encompassing M33's present-day morphology, its SFH, and the existence of the \HI\ bridge between these galaxies.\\

In contrast to these observations, we have shown that M31's Giant Stellar Stream cannot be reproduced with a hydrodynamical model for the infall of M33 onto M31 following an orbit with a close encounter in the remote past. This is consistent with the idea that the GSS is a structure created some $\sim 1 - 3$ Gyr ago by the direct collision of M31 with a companion featuring a range of possible masses \citep[$\sim 10^9 - 10^{11}$ \Msun; ][]{far06a,ham18a}, and which we have not included in our model. Therefore, the existence of the GSS and the plausibility of a close encounter between M31 and M33 in the distant past are not mutually exclusive.

A corollary of our hydrodynamical model is the existence of a yet undetected gas stream, the `Triangulum stream', that trails M33 as it falls towards M31. Even though this gas structure has not been directly observed so far, a possibly associated, high-density cloud may already be discernible in the 3D \HI\ maps of gas in and around M33 presented by \citet[][see also \citealt{put09a}]{kee16a}. We can think of a number of reasons why, if real, this stream has not been detected yet. Our model suggests that the stream is very diffuse with total gas column densities $\Ng \lesssim 10^{18}$ \psc. In addition, the gas along the stream could be highly ionised, and its distribution could be clumpy \citep[][]{bla17a}. Therefore, the stream may have a neutral gas fraction that is too low or too sparsely distributed to be observable with current radio facilities as a result of their limited sensitivity.

Largely dependent upon ionisation fraction, deep observations with FAST, such as the GAS, could potentially reach the required sensitivity to look for the presence of the Triangulum stream. It is worth noting that not even SKA1 will be as sensitive to extended \HI\ emission in the Local Volume. Other potential probes include targeting QSOs behind the region around M33, imaging \Ha\ beyond its optical disc \citep[][ see also \citealt{zhe17a} ]{kam15a}, and X-ray spectroscopy.

We conclude by pointing out some potential shortcoming of our hydrodynamic model, and suggesting avenues for improvement. First, we did not explore the role of different initial relative inclination of the galaxies play in the final configuration of the system. M33 is observed to lie roughly along the minor axis of M31 \citep[see][their figure 1]{leh15a}. This configuration is not faithfully reproduced by our model. However, it should not be difficult to come up with an initial relative orientation that broadly agrees with the observed one. Secondly, the IGM in our simulation is perhaps too diffuse compared to the Local Group IGM. In consequence, we may have underestimated the ram pressure exerted by the intra-group medium onto the gas streams which results from the motion of the system's barycentre through the IGM. Finally, we did not perform a test for the convergence of our simulation results. However, in \citet[][]{tep19a} we demonstrated that the limiting resolution we have adopted here is enough to model the structure of large-scale gas streams around galaxies. Nonetheless, in order to properly model the small-scale distribution of gas in the streams -- in particular the Triangulum stream, may require a higher resolution. That said, we expect all of these to have an effect only on the {\em detailed} structure of the tidal debris, but not on their large-scale features. These model improvements, in addition to detailed ionisation calculations necessary to quantify the \HI\ content of the gas distribution in and around the M31/M33 system, are all left for future work.

%--------------------------------------------------------------------------------------------------------------------------------------------------------------------------------
\section*{Acknowledgments}
We thank the referee for carefully reading our manuscript and providing a comprehensive and fair report.
This work is partially supported by the National Natural Science Foundation of China grant No. 11988101.
TTG acknowledges financial support from the Australian Research Council (ARC) through an Australian Laureate Fellowship awarded to JBH. 
We acknowledge access to the high-performance computing facility Raijin and additional resources on the National Computational Infrastructure (NCI),  which is supported by the Australian Government, through the University of Sydney's Grand Challenge Program the {\em Astrophysics Grand Challenge: From Large to Small} (CIs: Geraint F.~Lewis and JBH).
All figures and movie frames created with {\sc Gnuplot}, originally written by Thomas Williams and Colin Kelley.\footnote{\url{http://www.gnuplot.info} }
All animations assembled with {\sc ffmpeg}, \footnote{\url{http://www.ffmpeg.org}} created by Fabrice Bellard.
This research has made use of NASA's Astrophysics Data System (ADS) Bibliographic Services\footnote{\url{http://adsabs.harvard.edu} }, as well as of {\sc Pynbody} \footnote{\url{https://github.com/pynbody/pynbody}} and of {\sc Astropy} \citepalias{ast13a,ast18a},\footnote{\url{http://www.astropy.org} } both community-developed {\sc Python}\footnote{\url{http://www.python.org}, originally created by Guido van Rossum. } packages for Astronomy and analysis of simulation data, respectively.\\

%--------------------------------------------------------------------------------------------------------------------------------------------------------------------------------
% For the submission (both to journal and arXiv) comment out the 2 following lines,
% and un-comment the third:
%\bibliographystyle{mnras} % style mnras.bst
%\bibliography{/Users/tepper/references/complete} % references list
\input{bibliography.bbl} % use this when submitting both to journal and arXiv; be sure to include the .bbl file!

%--------------------------------------------------------------------------------------------------------------------------------------------------------------------------------
% Don't change these lines
\bsp	% typesetting comment
\label{lastpage}
\end{document}